\documentclass[journal]{IEEEtran}

\usepackage{amsmath} 
\usepackage{amssymb}  

\usepackage{color}
\usepackage{graphicx} 
\usepackage{aligned-overset}
\usepackage{subcaption}
\usepackage{cite}
\usepackage{dsfont}
\usepackage{mathtools}
\usepackage[shortlabels]{enumitem}

\newtheorem{thm}{\bf{Theorem}}[section]
\newtheorem{proposition}{\bf{Proposition}}[section]
\newtheorem{rem}{\bf{Remark}}[section]
\newtheorem{cor}{\bf{Corollary}}[section]
\newtheorem{defn}{\bf{Definition}}[section]

\newcommand{\nn}{\nonumber}

\newcommand{\vmm}{\mathbf{m}}

\newcommand{\vxx}{\mathbf{x}}

\newcommand{\vuu}{\mathbf{u}}

\newcommand{\vC}{\mathbf{C}}

\allowdisplaybreaks

\author{
Ertan~Kaz{\i}kl{\i}, Serkan~Sar{\i}ta\c{s},  Sinan~Gezici, and~Serdar~Y{\"u}ksel \thanks{E. Kaz{\i}kl{\i}, and S. Y\"uksel are with the Department of Mathematics and Statistics, Queen's University, K7L 3N6, Kingston, Ontario, Canada. Emails: ertan.kazikli@queensu.ca and yuksel@mast.queensu.ca. S. Sar{\i}ta\c{s} is with the Department of Electrical and Electronics Engineering, Middle East Technical University (METU), 06800, Ankara, Turkey. Email: ssaritas@metu.edu.tr. S. Gezici is with the Department of Electrical and Electronics Engineering, Bilkent University, 06800, Ankara, Turkey, Email: gezici@ee.bilkent.edu.tr.}
}

\title{Quadratic Signaling with Prior Mismatch at an Encoder and Decoder: Equilibria, Continuity and Robustness Properties}

\begin{document}

\maketitle

\begin{abstract}
We consider communications through a Gaussian noise channel between an encoder and a decoder which have subjective probabilistic models on the source distribution. Although they consider the same cost function, the induced expected costs are misaligned due to their prior mismatch, which requires a game theoretic approach. We consider two approaches: a Nash setup, with no prior commitment, and a Stackelberg solution concept, where the encoder is committed to a given announced policy apriori. We show that the Stackelberg equilibrium cost of the encoder is upper semi continuous, under the Wasserstein metric, as encoder's prior approaches the decoder's prior, and it is also lower semi continuous with Gaussian priors. For the Stackelberg setup, the optimality of affine policies for Gaussian signaling no longer holds under prior mismatch, and thus team-theoretic optimality of linear/affine policies are not robust to perturbations. We provide conditions under which there exist informative Nash and Stackelberg equilibria with affine policies. Finally, we show existence of fully informative Nash and Stackelberg equilibria for the cheap talk problem under an absolute continuity condition.
\end{abstract}


\section{Introduction}
The team theoretic formulation in systems theory (e.g., as studied by Witsenhausen \cite{WitsenhausenIntrinsic}) requires that all decision makers have the same probabilistic system model, even though they may have different local information. While this also has been the norm for nearly all information theoretic applications, in some applications, an encoder and a decoder may have subjective probabilistic models, especially when an encoder may realize that the model as seen by a remote decoder is inaccurate. Even though the decision makers employ the same cost function, induced expected costs, given encoding and decoding functions, are different from the perspective of the encoder and decoder due to their subjective probabilistic beliefs, which turns the team problem into a game theoretic one. For cooperative setups, the encoder needs to account for this inconsistency, which leads to a leader-follower (Stackelberg) game formulation. In some further applications, the encoder and the decoder may be engaged in a signaling game where their models may not be available to each other apriori or they may belong to separate organizations, leading to a Nash game theoretic setup. Accordingly, in this paper, we study communications between an encoder and a decoder, viewed as two decision makers, which have subjective beliefs on the probabilistic model of the source distribution. Depending on the cooperation or commitment nature of the encoder to its policies, we study Stackelberg and Nash equilibria for signaling problems and establish equilibrium solutions and their properties. These equilibria can be used to model different practical communication scenarios.

In scenarios modeled by the Stackelberg equilibrium concept, there is a hierarchy in the decision making procedure \cite{basols99}. In particular, the encoder first makes a decision and announces its decision and then the decoder acts after observing the encoder's decision. In this setting, the encoder commits to employ this announced strategy, and the decoder trusts the encoder and employs its best response. In this scenario, the encoder knows its own prior as well as the prior of the decoder whereas the decoder only knows its own prior. This happens especially when the encoder realizes that the decoder employs an inaccurate prior. In this setting, the encoder decides on what information to reveal to the decoder in order to optimize its objective function. This scenario can be viewed as a cooperative communication scenario since we know that the encoder's announcement is observed by the decoder and the decoder acts by using this information. We note that the classical communication setup with no strategic decision makers corresponds to the Stackelberg equilibrium concept since the decision makers trust each other in such a setting. Different from the classical communication setup, we incorporate prior mismatch into the problem. This type of cooperative communication setup with mismatched priors can be used to model scenarios where it is not feasible for the encoder to share its prior probability distribution with the decoder, the encoder conveys only the message, and the encoding function used for generating this message (which is designed off-line).

On the other hand, for a Nash equilibrium, there is no hierarchy in the decision making procedure and there is no commitment assumption \cite{basols99}. This happens for instance when decision makers do not trust announcements of each other, and thus keep in mind that the other decision maker may backtrack its commitment. This type of interaction is covered by the Nash equilibrium concept where each decision maker announces their policies at the same time. In this scenario, the decision makers do not need to know the prior distribution seen by the other decision maker. This scenario is also referred to as non-cooperative scenario since none of the decision makers take the other decision maker's announcement for granted.

\subsection{Preliminaries}

\begin{figure}
\centering
\null\hfill
\begin{subfigure}[t]{0.48\textwidth}
\centering
\includegraphics[scale=0.9]{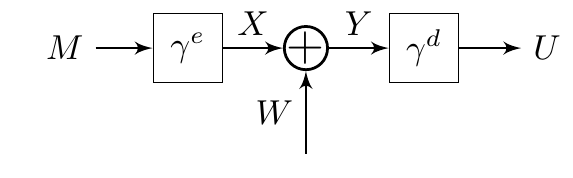}
\caption{Signaling game.}
\label{figure:genScheme}
\end{subfigure}
\hfill
\begin{subfigure}[t]{0.48\textwidth}
\centering
\includegraphics[scale=0.9]{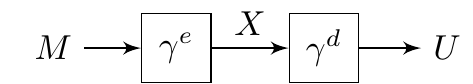}
\caption{Cheap talk.}
\label{figure:noiselessScheme}
\end{subfigure}
\hfill\null
\caption{Signaling game models.}
\label{figure:allScheme}
\end{figure}

A {\it signaling game} problem, which is depicted in Fig.~\ref{figure:allScheme}-(\subref{figure:genScheme}), can be formulated as follows: Suppose that there is an encoder and a decoder. The encoder wishes to transmit a random variable $M$ taking values in $\mathbb{M}$. The encoder encodes $M$ into an $\mathbb{X}$-valued random variable $X$ via an encoding policy denoted by $\gamma^e \in \Gamma^e$ where $\Gamma^e$ denotes the set of possible encoding strategies. We take $\Gamma^e$ to be the set of all stochastic kernels from $\mathbb{M}$ to $\mathbb{X}$.\footnote{$P$ is a stochastic kernel from $\mathbb{M}$ to $\mathbb{X}$ if $P(\cdot|m)$ is a probability measure on ${\mathcal B}(\mathbb{X})$ for every $m\in \mathbb{M}$, and $P(A|\cdot)$ is a Borel measurable function of $m$ for every $A\in{\mathcal B}(\mathbb{X})$. Note that if $\mathbb{M}$ and $\mathbb{X}$ are finite, then $P$ corresponds to a transition matrix.} There exists an additive Gaussian noise channel between the encoder and decoder where the additive noise is independent of the source. In this setup, the encoder designs its policy by incorporating either a soft or a hard power constraint into its objective function, as explained below. The decoder observes a noise corrupted version of the message denoted by $Y=X+W$ which is a $\mathbb{Y}$-valued random variable. The decoder generates its optimal decision $U$, which is also an $\mathbb{M}$-valued random variable, given its observation $Y$ via a decoding policy $\gamma^d \in \Gamma^d$. Here, $\Gamma^d$ denotes the set of possible decoding strategies which is the set of all stochastic kernels from $\mathbb{Y}$ to $\mathbb{M}$. We also consider a {\it cheap talk} setup where the encoded message is directly observable by the decoder and there is no power constraint at the encoder and this setup is depicted in Fig.~\ref{figure:allScheme}-(\subref{figure:noiselessScheme}).

In many applications (in networked systems, recommendation systems, and applications in economics) the objectives of the encoder and the decoder, and the perception on the probability measure of the common source may not be aligned. For example, the encoder may aim to minimize
$
J^e(\gamma^e,\gamma^d) = \mathbb{E}_e\left[c^e(m, u)\right]
$
whereas the decoder may aim to minimize
$
J^d(\gamma^e,\gamma^d) = \mathbb{E}_d\left[c^d(m, u)\right]
$
where $c^e(m,u)$ and $c^d(m,u)$ denote the cost functions of the encoder and the decoder, respectively, when the action $u$ is taken for the corresponding message $m$, and $\mathbb{E}_e[\cdot]$ and $\mathbb{E}_d[\cdot]$ denote that the expectation is taken over the probability measure from the perspective of the encoder and the decoder, respectively.\footnote{When we need to emphasize that we are working with a random variable, we consider upper case letters, otherwise we use small letters both for realizations and the variables and emphasize their distinction when there is room for confusion.} Note that each decision maker computes their expected cost with respect to its own subjective prior since it believes that its subjective prior is the true prior distribution. Each player designs its policy by minimizing its expected cost computed with respect to its subjective prior distribution. Thus, although the actual true distribution can be a different distribution than these subjective prior distributions, this true distribution does not affect policies employed by the decision makers. Nevertheless, in the following remark, we provide a motivation for a case that the encoder's prior is the true prior distribution. 

\begin{rem}
Note that the encoder designs its message by observing the source random variable. Thus, in some applications, it may be possible for the encoder to correct its prior before the information transmission stage by observing a large number of samples. Therefore, in these applications, it is reasonable to assume that the encoder's view corresponds to the true distribution of the source. On the other hand, the decoder may have an incorrect belief since its observations are limited to what the encoder reveals. 
\end{rem}

In this paper, we consider a quadratic cost structure where either a soft power constraint or a hard power constraint is employed at the encoder. In the case of soft power constraint, the encoder employs the following objective function
\begin{align} \label{eq:JeSoft}
J^e(\gamma^e,\gamma^d) = \mathbb{E}_e\left[c^e(m,x,u)\right],
\end{align}
and the decoder employs the objective function
\begin{align} \label{eq:JdSoft}
J^d(\gamma^e,\gamma^d) = \mathbb{E}_d\left[c^d(m,u)\right],
\end{align}
where $c^e(m,x,u) = (m-u)^2+\lambda x^2$ and $c^d(m,x) = (m-u)^2$ and $\lambda$ represents the appended soft power constraint to encoder's objective. Appending a soft power constraint in this manner is encountered in stochastic control problems, see e.g., \cite{witsenhausen1968counterexample,BasarCDC2008}. Notice that when $\lambda =0$, the case without any power constraint is recovered. Note also that it is possible append this additional $\lambda x^2$ term to the cost function of the decoder and this does not make a difference in the analysis. In other words, misalignment between the encoder and decoder essentially arises from subjective probabilistic beliefs of the players and not from the considered costs. For the case in which the encoder has a hard power constraint instead of a soft power constraint, the goal of the encoder is to minimize 
\begin{align} \label{eq:JeHard}
J^e(\gamma^e,\gamma^d) &= \mathbb{E}_e\left[c^e(m,u)\right]\nn\\
& \text{s.t. } \mathbb{E}_e\left[\left(\gamma^e(m)\right)^2\right] \leq \overline{P},
\end{align}
whereas the decoder aims to minimize
\begin{align} \label{eq:JdHard}
	J^d(\gamma^e,\gamma^d) = \mathbb{E}_d\left[c^d(m,u)\right], 
\end{align}
where $c^e\left(m,u\right) = \left(m-u\right)^2$ and $c^d\left(m,u\right) = \left(m-u\right)^2$. In communication theoretic settings, a hard power constraint for an encoder is commonly imposed, and many results in information theory with regard to communication through a Gaussian channel involves a hard power constrained encoder, see e.g., \cite[Ch.~9]{CoverThomasBook} and \cite[Ch.~11]{YukselBasarBook}.

Our aim is to analyze the previously described communication scenarios using two important game theoretic concepts which make different assumptions on how decision makers interact: Nash equilibrium and Stackelberg equilibrium \cite{basols99}. Under the Nash equilibrium concept, the encoder and the decoder announce their policies simultaneously without knowing the policy of the other player or with no commitment to the announced plays. A set of policies is said to be a Nash equilibrium if neither of the players has incentive to unilaterally deviate from its current strategy. In particular, a pair of encoding and decoding policies $\gamma^{*,e}$ and $\gamma^{*,d}$ forms a Nash equilibrium if \cite{basols99}
\begin{align}
J^e(\gamma^{*,e}, \gamma^{*,d}) &\leq J^e(\gamma^{e}, \gamma^{*,d}) \quad \forall \gamma^e \in \Gamma^e ,\\
J^d(\gamma^{*,e}, \gamma^{*,d}) &\leq J^d(\gamma^{*,e}, \gamma^{d}) \quad \forall \gamma^d \in \Gamma^d .
\end{align}
We note that in this paper we consider only subjective Gaussian priors for the Nash equilibria analysis.

On the other hand, under the Stackelberg equilibrium concept, the game is played in a sequential manner where first the encoder chooses and announces its policy and then the decoder determines its policy given the announcement of the encoder. In this scenario, the encoder commits to employ its announced strategy and the encoder cannot change its strategy once the decoder learns the strategy of the encoder. Since the decoder knows the strategy of the encoder, it takes its optimal response given the announced strategy of the encoder. A pair of encoding and decoding policies $\gamma^{*,e}$ and $\gamma^{*,d}$ forms a Stackelberg equilibrium if \cite{basols99}
\begin{align}
&J^e(\gamma^{*,e}, \gamma^{*,d}(\gamma^{*,e})) \leq J^e(\gamma^e, \gamma^{*,d}(\gamma^e)) \quad \forall \gamma^e \in \Gamma^e ,\\
&\hspace{-0.5cm} \text{where } \gamma^{*,d}(\gamma^e) \text{ satisfies} \nonumber \\
&J^d(\gamma^{e}, \gamma^{*,d}(\gamma^{e})) \leq J^d(\gamma^{e}, \gamma^d(\gamma^{e})) \quad \forall \gamma^d \in \Gamma^d,
\label{eq:stackelbergEquilibrium}
\end{align}
where for the policy of the decoder we use the notation $\gamma^{d}(\gamma^e)$ to indicate that the decoder decides on its policy after observing the encoder's policy. In contrast to Nash equilibria, we present general results concerning arbitrary distributions for Stackelberg equilibria analysis with more specific results regarding the Gaussian case.

\begin{rem}
Note that, under the Stackelberg assumption, the encoder must know decoder's subjective prior so that it, as a leader, can anticipate decoder's optimal actions. On the other hand, for the Nash case, the agents do not need to know their subjective priors; they know only their policies as they (simultaneously) announce to each other. 
\end{rem}

In a signaling game problem, it is of interest to investigate existence of equilibrium in which the encoder conveys information related to its observation, i.e., the encoded random variable $X$ is not statistically independent of the source random variable $M$. Such kind of equilibria are referred to as {\it informative equilibrium}. On the other hand, if the encoded random variable does not depend on the source $M$, this type of equilibrium is referred to as {\it non-informative equilibrium} or {\it babbling equilibrium}.

For the Stackelberg setup, an important quantity is the induced cost for the encoder as the encoder performs an optimization of its objective given that the decoder best responds. In that respect, we investigate the continuity properties of the encoder's cost with respect to perturbations of the priors around the team setup. We say that a Stackelberg equilibrium is {\it robust} with respect to perturbations around the team setup if the encoder's expected cost continuously changes as the prior of the encoder (decoder) approaches the prior of the decoder (encoder). We also have results where one can only guarantee upper semi continuity or lower semi continuity of the encoder's cost. Moreover, we also investigate continuity properties for the affine Nash equilibria. 

For the Stackelberg setup, in the case when the cost is upper semi continuous, there cannot be a drastic degradation in encoder's performance with prior mismatch around the point of identical priors. In that respect, upper semi continuity ensures that the worst case performance under prior mismatch behaves continuously. On the other hand, in the case when the cost is only lower semi continuous, a small prior perturbation in principle may lead to a large performance degradation of the game theoretic cost in comparison with the team theoretic cost.

\subsection{Literature Review}

Crawford and Sobel in their seminal work \cite{SignalingGames} investigate a communication scenario between an encoder and a decoder which do not share a common objective function due to a bias term appearing in the objective function of the encoder. They establish that under certain technical conditions the encoder is required to apply a quantization policy at a Nash equilibrium. In particular, due to the misalignment in the objective functions of the encoder and the decoder, the encoder hides information by reporting the quantization bin that its observation lies in, rather than revealing its observation completely. This is in contrast with the classical team theoretic communication setup where revealing more information is always beneficial for the system. In contrast with the Nash setup, it is also possible to consider a Stackelberg game setup and an important line of work in this context in the economics literature is the Bayesian persuasion problem where signaling scenarios are investigated under the Stackelberg equilibrium concept \cite{bayesianPersuasion}.

Signaling game problems find applications in various contexts including communication and control theory literature \cite{EAkyolProcIEEE2017,tacQuadraticSignalingSaritas,EstStrategicSensorsFarokhi2017,Tamura2018,SaritasAutomatica2020,MOSayinAutomatica2019,TreustPersuasion2019,treust2018persuasion,ProspectBasar2018,SaritasISIT2019,VoraCDC2020,VoraISIT2020}. For instance, the work in \cite{EAkyolProcIEEE2017} considers a quadratic cost structure for transmitting a scalar Gaussian source from an encoder to a decoder where encoder's cost includes a bias term which is modeled as jointly Gaussian with the source message. The authors analyze such communication scenarios under the Stackelberg equilibrium concept and derive equilibrium solutions, which turn out to be linear. In \cite{Tamura2018}, signaling scenarios under a general quadratic cost structure are investigated under the Stackelberg equilibrium concept and for multidimensional Gaussian sources, the optimality of linear policies is established for the considered cost structure. The work in \cite{tacQuadraticSignalingSaritas} investigates communication scenarios between an encoder and a decoder under quadratic costs using either Nash or Stackelberg equilibria concepts where the encoder's objective contains a deterministic bias term. An important observation from \cite{tacQuadraticSignalingSaritas} is the existence of linear Nash equilibria for Gaussian sources. We note that various studies consider also more general cost functions \cite{bayesianPersuasionHetPriors,TreustPersuasion2019,treust2018persuasion,VoraCDC2020,VoraISIT2020}, rather than focusing on the quadratic case. For instance, \cite{bayesianPersuasionHetPriors} analyzes the Bayesian persuasion problem with general cost functions where the encoder and decoder have subjective probabilistic beliefs. In addition, the works \cite{TreustPersuasion2019,treust2018persuasion} analyze information theoretic limits of the Bayesian persuasion problem with general cost functions. Another related work \cite{CGokenTWC2010} investigates optimal stochastic signaling in a binary communication setup with aligned cost structure for an encoder and a decoder (i.e., team theoretic setup) and provides sufficient conditions under which stochastic signaling improves the performance or not. 

Subjective probabilistic models are encountered in various contexts. For instance, the setup in decentralized decision making where the priors of decision makers may be different has a practical significance. There have been a number of studies on the presence of a mismatches in the priors of decision makers \cite{BasTAC85, TeneketzisVaraiya88, CastanonTeneketzis88,bayesianPersuasionHetPriors,SaritasTSP2019}. In such setups, even when the objective functions to be optimized are identical, the presence of subjective priors alters the formulation from a team problem to a game problem involving strategy/policy spaces (see \cite[Section 12.2.3]{YukselBasarBook} for a comprehensive literature review on subjective priors also from a statistical decision making perspective). For example, \cite{BasTAC85} investigates equilibrium behavior under either Nash or Stackelberg equilibria in a two-person decision making scenario with quadratic cost where the decision makers have subjective probabilistic beliefs. An interesting observation from \cite{BasTAC85} is that in the special case of Gaussian priors, the decision makers employ linear policies under Nash equilibria whereas the policies under Stackelberg equilibria are in general nonlinear. In addition, in almost all practical applications, there is some mismatch between the true and an assumed probabilistic system/data model, which results in performance degradation. This performance loss due to the presence of mismatch has been studied extensively in various setups (see e.g., \cite{mismatchedEstimation,estimationRobustnessMismatch,mismatchSurvey,Gray1975,MismatchedMMSEISIT2012,MismatchGuessworkITW2019,JubaCompression2011,BravermanIT2019} and references therein). Moreover, the subjectivity may appear when there are prospect theoretic agents in the system where the decision makers may have different views on the probabilistic models due to their subjective biases \cite{ProspectBasar2018,GeziciProspect2018,ProspectVarshney2020}. In prospect theory, the subjective views of the agents on the prior probabilities are modeled via a weight function which for instance reflects a common misconception of overestimating (underestimating) the probability of less (more) likely events. For instance, in \cite{ProspectBasar2018}, communication scenarios through an additive noisy channel with an encoder and a decoder which have different weight functions is analyzed under the Stackelberg equilibrium concept where there is an affine policy restriction at the encoder. It is shown that for Gaussian source and noise case policies at the equilibrium are not affected by subjective biases whereas for exponential source and noise case policies at the equilibrium depend on subjective biases.

\subsection{Contributions} 
We analyze the signaling game problem described earlier under the Stackelberg equilibrium or the Nash equilibrium concepts. Our results concerning the Stackelberg equilibria involve arbitrary distributions with more specific results for the Gaussian case. On the other hand, we focus on the Gaussian case for the Nash equilibria analysis. In addition, we consider a cheap talk problem under the Stackelberg or Nash equilibria for which mutually absolutely continuity assumption\footnote{Both subjective probability measures agree on the sets with zero measure, i.e., the Radon-Nikodym derivative of either measure with respect to the other exists.} is made for the subjective prior distributions. Main contributions of this manuscript can be summarized as follows:
\begin{enumerate}[(i)]
\item We prove that the Stackelberg equilibrium cost of the encoder is upper semi continuous (under the Wasserstein metric) when the prior of encoder is perturbed from the prior of decoder considering any subjective prior distributions for the players (Theorem~\ref{thm:robustness}). For the special case of Gaussian priors, it is proven that the cost around the team setup is lower semi continuous, as well. Therefore, for the Gaussian priors case, the equilibrium is robust with respect to perturbations around the team setup (in both the Wasserstein metric and under weak convergence) where robustness refers to the fact that the Stackelberg equilibrium cost of the encoder is continuous with respect to prior perturbation around the team setup. In addition, we also provide a duality result which states that if the prior of decoder is perturbed from the prior of encoder, the Stackelberg equilibrium cost of encoder is lower semi continuous (under both the Wasserstein metric or the weak convergence topology).

\item We show that the Stackelberg equilibrium solution for Gaussian signaling with subjective priors is in general nonlinear by providing specific examples where a nonlinear policy outperforms the best affine policy (Theorem~\ref{thm:StackelbergSoftQuant}). Thus, team-theoretic optimality of linear/affine policies are not robust to prior perturbations.

\item For the signaling game problem with Gaussian priors (under a soft power constraint or a hard power constraint), we show that the Stackelberg equilibrium under affine policy restriction is either informative or non-informative depending on conditions stated explicitly (Theorem~\ref{thm:StackelbergSoft} and Theorem~\ref{thm:StackelbergHard}). Moreover, these game theoretic solutions do not coincide with the corresponding team theoretic solutions in general. In particular, when there is a hard power constraint, game theoretic solution may be non-informative whereas team theoretic solution is always informative regardless of the parameters. For the signaling game problem with Gaussian priors under a soft power constraint, we show that there exists a unique informative affine Nash equilibrium under a certain condition involving the subjective prior of the decoder (Theorem~\ref{thm:NashScalarSoft}). On the other hand, we prove that there always exists a unique informative affine Nash equilibrium when there is a hard power constraint at the encoder (Theorem~\ref{thm:NashHard}). In addition, we show that the informative affine Nash equilibrium solution under a soft power constraint or a hard power constraint is robust to perturbations around the team setup.

\item We show that there exist fully informative Nash and Stackelberg equilibria for the dynamic cheap talk as in the team theoretic setup when the encoder and the decoder have subjective priors on the source distribution and identical costs, provided that the priors are mutually absolutely continuous (Theorem~\ref{thm:cheap}).
\end{enumerate}

\section{The Cooperative/Commitment Setup (Stackelberg Equilibria)}

In this section, we analyze the Stackelberg equilibria when there is either a soft power constraint or a hard power constraint at the encoder. Before presenting our results, we make the following remark.

\begin{rem}
We note that if both players share the same probabilistic belief on the source distribution, then the problem reduces to the classical team theoretic setup with a power constrained encoder. Although obtaining optimal coding/decoding policies for general source distributions is in general difficult, for the special case of scalar Gaussian source, the optimal solution involves linear policies at the encoder and decoder, see e.g., \cite[p.~376]{YukselBasarBook}. This optimality result for linear policies is obtained by using channel capacity and rate distortion bounds. 
\end{rem}

\subsection{Continuity and Robustness to Perturbations around the Team Setup}

In this subsection, we investigate continuity and robustness of the Stackelberg equilibria around the team setup for general subjective source priors with more specific results for the Gaussian priors case. In the literature, analytical properties such as continuity of mean squared error for estimation under additive noise are investigated in \cite{WuVerdu2012} when there is no prior mismatch, i.e., a team theoretic setup. In addition, \cite{ADKara2019SIAM} investigates robustness and continuity with respect to prior probability measures for partially observed stochastic control problems where the priors can be incomplete or incorrect. Here, \cite{ADKara2019SIAM} shows that indeed under total variation, strong continuity results with a rate of continuity/convergence hold, but in our work, the presence of an encoder adds further challenges. We leave this problem for future work.

In our work, the subjective probabilistic belief of the one of the players deviates from that of the other player, i.e., a deviation from the team theoretic setup. In this case, we analyze if the cost function of the encoder behaves continuously with respect to such a difference on the priors. To be more precise, a robust Stackelberg equilibrium means that $ J^{*,e}(\phi_e,\phi_d) \to J^{*,e}(\phi_d,\phi_d)$ as $\phi_e\to\phi_d$, where $J^{*,e}(\phi_e,\phi_d)$ denotes the Stackelberg equilibrium cost of the encoder when the priors of the encoder and decoder are $\phi_e(\cdot)$ and $\phi_d(\cdot)$, respectively. Such a continuity result is equivalent to $J^{*,e}(\phi_e,\phi_d) \to J^{*,e}(\phi_e,\phi_e)$ as $\phi_d\to\phi_e$. We also have semi continuity results where depending on whether $\phi_e\to\phi_d$ or $\phi_d\to\phi_e$ is considered, upper or lower semi continuity can be guaranteed. 

While performing a continuity analysis with respect to subjective priors, one applies a perturbation to these subjective priors around the point of identical priors. This perturbation is quantified via a convergence notion for probability measures as defined in the following. We also emphasize that these continuity results hold only at the point of identical priors.

In order to investigate continuity properties of the encoder's cost, we need to define a probability space and a convergence notion for probability measures in this space. Towards that goal, let $\mathbb{X}=\mathbb{R}$ and let $\mathcal{P}(\mathbb{X})$ denote the family of all probability measures on $(\mathbb{X},\mathcal{B}(\mathbb{X}))$ where $\mathcal{B}(\mathbb{X})$ denotes the Borel $\sigma$-algebra on $\mathbb{X}$. Let $\{\mu_n,\, n\in \mathbb{N}\}$ be a sequence in $\mathcal{P}(\mathbb{X})$. A sequence $\{\mu_n\}$ is said to converge to $\mu\in \mathcal{P}(\mathbb{X})$ as $n$ tends to infinity \emph{weakly} if the following convergence relation holds as $n$ tends to infinity:
\[
 \int_{\mathbb{R}} c(x) \mu_n(dx)  \to \int_{\mathbb{R}}c(x) \mu(dx)
\]
for every continuous and bounded $c: \mathbb{X} \to \mathbb{R}$.

The Prohorov metric can be used to metrize this space. As a more practical metric, the Wasserstein metric can also be used (for compact $\mathbb{X}$). 
\begin{defn}[Wasserstein metric]
The \emph{Wasserstein metric} of order $p \geq 1$ for two distributions $\mu,\nu\in\mathcal{P}(\mathbb{X})$ with finite $p$th order moments is defined as
\[  W_p(\mu,\nu) = \inf_{\eta \in \mathcal{H}(\mu,\nu)} \bigg( \int_{\mathbb{X}\times\mathbb{X}} \eta(dx,dy) \|x-y\|^p\bigg)^{\frac{1}{p}},\]
where $\mathcal{H}(\mu,\nu)$ denotes the set of probability measures on $\mathbb{X}\times\mathbb{X}$ with first marginal $\mu$ and second marginal $\nu$ and $\|\cdot\|$ is a norm (such as the Euclidean norm).
\end{defn}

As noted, for compact $\mathbb{X}$, the Wasserstein distance of order $p$ metrizes the weak topology on the set of probability measures on $\mathbb{X}$ (see \cite[Theorem~6.9]{villani2009optimal}). For non-compact $\mathbb{X}$, weak convergence combined with convergence of moments (that is of $\int \mu_n(dx) \|x\|^q \to \int \mu(dx) \|x\|^q$ for all orders $q\leq p$) is equivalent to convergence in $W_p$ (see \cite[Definition~6.8]{villani2009optimal} and \cite[Theorem~6.9]{villani2009optimal}).

In the following theorem, we investigate continuity and robustness properties of the encoder's cost around the point of identical priors. We analyze the cost of the encoder since in the Stackelberg setup the encoder performs an optimization of its objective, and thus, encoder's objective determines whether the equilibrium is robust or not.

\begin{thm}\label{thm:robustness}
Suppose that the prior of the source is $\phi_e(\cdot)$ and $\phi_d(\cdot)$ from the perspective of the encoder and the decoder, respectively, where these prior distributions are arbitrary. Suppose further that variance of the source under $\phi_d(\cdot)$ is finite. Then, the following are true where there is a either a soft power constraint or a hard power constraint at the encoder.
\begin{enumerate}[(i)]
\item The Stackelberg equilibrium cost of the encoder is upper semi continuous (under the Wasserstein metric) as subjective prior of the encoder approaches subjective prior of the decoder, i.e., $\phi_e(\cdot)\to\phi_d(\cdot)$.
\item If the prior distributions $\phi_e(\cdot)$ and $\phi_d(\cdot)$ are Gaussian, the Stackelberg equilibrium cost of the encoder is lower semi continuous as subjective prior of the encoder approaches subjective prior of the decoder, i.e., $\phi_e(\cdot)\to\phi_d(\cdot)$.
\item If the prior distributions $\phi_e(\cdot)$ and $\phi_d(\cdot)$ are Gaussian, the Stackelberg equilibria are robust with respect to perturbations around the team setup (under both the Wasserstein metric or the weak convergence topology), i.e., the Stackelberg equilibrium cost of the encoder is continuous.
\item The Stackelberg equilibria cost of the encoder is lower semi continuous (under both the Wasserstein metric or the weak convergence topology) as subjective prior of the decoder approaches subjective prior of the encoder, i.e., $\phi_d(\cdot)\to\phi_e(\cdot)$.
\end{enumerate}
\end{thm}

\begin{figure}
\centering
\includegraphics[scale=0.9]{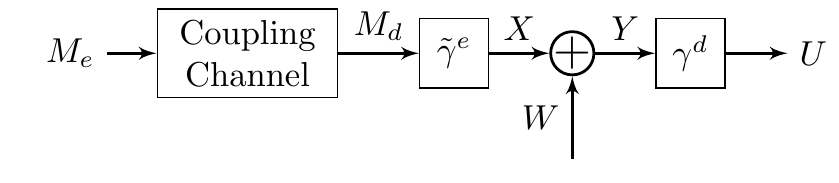}
\caption{Suboptimal scheme used in proving upper semi continuity result where the coupling channel $f(\cdot)$ is such that marginal distribution of its output is $\phi_d(\cdot)$.}
\label{fig:suboptimal}
\end{figure}

\begin{rem}
For a communication setup, it is reasonable to assume that the source to be conveyed has a finite average power, which holds when the variance of the source is finite. Please also note that if the variance is unbounded, infinite Shannon capacity would be implied, which would reduce the problem to a cheap talk problem, as the noise can be suppressed with arbitrarily large encoder gain. Therefore, the assumption of having a finite variance from the perspective of the decoder in Theorem~\ref{thm:robustness} is not unnatural, and this assumption is commonly imposed in practice.
\end{rem}

\begin{IEEEproof}
In the following, we focus on the case with a soft power constraint. We note that all the derived identities in the following proof identically hold when the soft power constraint is replaced with a hard power constraint. Therefore, the proof in the case of a hard power constraint follows the same steps as in the proof for the case with a soft power constraint.
\begin{enumerate}[(i)]
\item Let $M_e$ and $M_d$ denote random variables distributed according to the priors $\phi_e(\cdot)$ and $\phi_d(\cdot)$, respectively. The Stackelberg equilibrium cost is given by
\begin{align}
\min_{\gamma^e(\cdot)} \mathbb{E}[(m_e-u)^2] + \lambda \mathbb{E}[\gamma^e(m_e)^2]
\end{align}
where there is an implicit constraint that the decoder employs its best response with respect to its own prior. In particular, the decoder's action is a deterministic function of its observation given by $u=\mathbb{E}_d[m|\gamma^e(m)+w]$. Now consider a suboptimal scheme in which the encoder maps its observation into an auxiliary variable via $f(\cdot)$ which is then mapped into the transmitted message through $\tilde{\gamma}^e(\cdot)$ under constraint that the marginal distribution of this auxiliary random variable is fixed to $\phi_d(\cdot)$. This scheme is depicted in Fig.~\ref{fig:suboptimal}. By introducing such an auxiliary random variable which has a fixed marginal distribution, the proposed scheme is in general a suboptimal scheme. Here, the encoder is given by $\gamma^e(m_e) = \tilde{\gamma}^e(f(m_e))$ where $f(m_e)\triangleq m_d$ has a fixed marginal distribution $\phi_d(\cdot)$. 
Observe that 
\begin{align}
&\min_{\gamma^e(\cdot)} \mathbb{E}[(m_e-u)^2] + \lambda \mathbb{E}[\gamma^e(m_e)^2]\nn \\
&\leq 
\min_{\tilde{\gamma}^e(\cdot),f(\cdot)} \mathbb{E}[(m_e-u)^2] + \lambda \mathbb{E}[\tilde{\gamma}^e(m_d)^2]\nn\\
&=
\min_{\tilde{\gamma}^e(\cdot),f(\cdot)} \mathbb{E}[(m_e-m_d+m_d-u)^2] +  \lambda \mathbb{E}[\tilde{\gamma}^e(m_d)^2]\nn\\
&=
\min_{\tilde{\gamma}^e(\cdot),f(\cdot)} \mathbb{E}[(m_e-m_d)^2] + \mathbb{E}[(m_d-u)^2] \nn\\
&\hphantom{=\min_{\tilde{\gamma}^e(\cdot),f(\cdot)}}
+2 \mathbb{E}[(m_e-m_d)(m_d-u)]+\lambda \mathbb{E}[\tilde{\gamma}^e(m_d)^2]\nn\\
&\leq 
\min_{\tilde{\gamma}^e(\cdot),f(\cdot)} \mathbb{E}[(m_e-m_d)^2] + \mathbb{E}[(m_d-u)^2] \nn\\
&\hphantom{=\min_{\tilde{\gamma}^e(\cdot),f(\cdot)}}+ 2\sqrt{\mathbb{E}[(m_e-m_d)^2]}\sqrt{ \mathbb{E}[(m_d-u)^2]} \nn \\
&\hphantom{=\min_{\tilde{\gamma}^e(\cdot),f(\cdot)}}
+\lambda \mathbb{E}[\tilde{\gamma}^e(m_d)^2] \label{eq:robustnessEq1}
\end{align}	
where the first inequality is due to suboptimality of the scheme with $f(\cdot)$ and $\tilde{\gamma}^e(\cdot)$ and the second inequality follows from Cauchy-Schwarz inequality. The final optimization problem in \eqref{eq:robustnessEq1} can be first solved with respect to $f(\cdot)$ which leads to 
\begin{align}
&\min_{\tilde{\gamma}^e(\cdot),f(\cdot)} \mathbb{E}[(m_e-m_d)^2] + \mathbb{E}[(m_d-u)^2]+\lambda \mathbb{E}[\tilde{\gamma}^e(m_d)^2] \nn\\
&\hphantom{\min_{\tilde{\gamma}^e(\cdot),f(\cdot)}}
+2\sqrt{\mathbb{E}[(m_e-m_d)^2]}\sqrt{ \mathbb{E}[(m_d-u)^2]} \nn\\
&=\min_{\tilde{\gamma}^e(\cdot)} W_2(\phi_e,\phi_d)^2+ \mathbb{E}[(m_d-u)^2] +\lambda \mathbb{E}[\tilde{\gamma}^e(m_d)^2]\nn\\
&\hphantom{=\min_{\tilde{\gamma}^e(\cdot)}}
+2\,W_2(\phi_e,\phi_d)\sqrt{ \mathbb{E}[(m_d-u)^2]} \label{eq:robustnessEq2}
\end{align}
where $W_2(\phi_e,\phi_d)$ denotes quadratic Wasserstein distance between the distributions $\phi_e(\cdot)$ and $\phi_d(\cdot)$. Note that the optimal value for the solution to \eqref{eq:robustnessEq2} is finite when $W_2(\phi_e,\phi_d)$ is finite. To see this, observe that if $\tilde{\gamma}^e(m_d)=0$ for all $m_d$, then the optimal decoder action becomes $u=\mu_d$ where $\mu_d$ denotes the mean of the source message from decoder's perspective. Then, the objective function in \eqref{eq:robustnessEq2} takes the value of $(W_2(\phi_e,\phi_d)+\mathbb{E}[(m_d-\mu_d)^2]^{1/2})^2$, which is finite as the variance of the source is finite from the perspective of decoder. Thus, an optimal solution to \eqref{eq:robustnessEq2} must yield a finite objective value. From this observation, it follows that as $\phi_e\to\phi_d$, the optimal solution of \eqref{eq:robustnessEq2} leads to a finite value for the term $\mathbb{E}[(m_d-u)^2]$ since $W_2(\phi_e,\phi_d)\to 0$ in this case. As a result, we get
\begin{align*}
&\limsup_{\phi_e\to\phi_d} \min_{\gamma^e(\cdot)} \mathbb{E}[(m_e-u)^2] + \lambda \mathbb{E}[\gamma^e(m_e)^2]\\
&\leq \limsup_{\phi_e\to\phi_d} \min_{\tilde{\gamma}^e(\cdot)} W_2(\phi_e,\phi_d)^2 + \mathbb{E}[(m_d-u)^2] \\
&\hphantom{PHA} 
+ 2\,W_2(\phi_e,\phi_d)\sqrt{ \mathbb{E}[(m_d-u)^2]} 
+\lambda \mathbb{E}[\tilde{\gamma}^e(m_d)^2]\\
&= \mathrm{E}[(m_d-\mathrm{E}[m_d|\gamma^{*,e}(m_d)+w])^2] +  \lambda \mathbb{E}[\gamma^{*,e}(m_d)^2]
\end{align*}
where the inequality is due to \eqref{eq:robustnessEq1} and \eqref{eq:robustnessEq2}, $\gamma^{*,e}(\cdot)$ denotes the optimal encoding policy under team theoretic setup with common prior distribution $\phi_d(\cdot)$ and the equality follows from the facts that $\mathbb{E}[(m_d-u)^2]$ is finite and $W_2(\phi_e,\phi_d)\to 0$ as $\phi_e\to\phi_d$. Since the last term corresponds to the cost for team theoretic setup with prior $\phi_d(\cdot)$, this analysis shows that the Stackelberg equilibrium cost is upper semi continuous around the team setup.
\item We note that if the decoder employs the encoder's prior rather than its own prior, the objective function of the encoder is improved. Therefore, the following inequality holds:
\begin{align}
&\min_{\gamma^e(\cdot)}\mathbb{E}_e[(m-\mathbb{E}_d[m|\gamma^e(m)+w])^2] + \lambda \mathbb{E}_e[x^2]\nn \\
&\;\geq \min_{\gamma^e(\cdot)}\mathbb{E}_e[(m-\mathbb{E}_e[m|\gamma^e(m)+w])^2] + \lambda \mathbb{E}_e[x^2]\label{eq:GaussianLowerBound}
\end{align}
For the latter optimization problem, it is well-known that the optimal encoding and decoding policies are affine. By using this observation, we get
\begin{align}
&\liminf_{\phi_e\to\phi_d} \min_{\gamma^e(\cdot)}\mathbb{E}_e[(m-\mathbb{E}_d[m|\gamma^e(m)+w])^2] + \lambda \mathbb{E}_e[x^2] \nn\\
&\geq \liminf_{\phi_e\to\phi_d} \min_{\gamma^e(\cdot)}\mathbb{E}_e[(m-\mathbb{E}_e[m|\gamma^e(m)+w])^2] + \lambda \mathbb{E}_e[x^2] \nn\\
&=\liminf_{\phi_e\to\phi_d} \mathbb{E}_e[(m-\beta_1^{*,d}(\alpha_1^{*,e} m+\alpha_2^{*,e}+w)-\beta_2^{*,d})^2] \nn \\
&\hphantom{=\liminf_{\phi_e\to\phi_d}} + \lambda \mathbb{E}_e[(\alpha_1^{*,e} m+\alpha_2^{*,e})^2] \nn\\
&= \mathbb{E}_d[(m-\beta_1^{*,d}(\alpha_1^{*,d} m+\alpha_2^{*,d} +w)-\beta_2^{*,d})^2] \nn\\
&\hphantom{=}+ \lambda \mathbb{E}_d[(\alpha_1^{*,d} m+\alpha_2^{*,d})^2] \label{eq:robustnessGaussian}
\end{align}
where $\alpha_1^{*,e}$, $\alpha_2^{*,e}$, $\beta_1^{*,e}$ and $\beta_2^{*,e}$ (resp. $\alpha_1^{*,d}$, $\alpha_2^{*,d}$, $\beta_1^{*,d}$ and $\beta_2^{*,d}$) are the optimal coefficients at the affine encoder and decoder for the problem of transmitting a Gaussian source $M_e$ (resp. $M_d$) over an independent additive Gaussian channel under quadratic criterion with soft power constraint. This establishes the lower semi continuity of the Stackelberg equilibrium cost around the point of identical priors. 

\item The result follows from (i) and (ii). 
\item The lower semi continuity result in this case can be established easily. In particular, we note that the inequality in \eqref{eq:GaussianLowerBound} holds for general subjective priors. In other words, if the decoder employs encoder's prior rather than its own prior while computing its estimate, the cost of the encoder improves. From this inequality, we get
\begin{align}
&\liminf_{\phi_d\to\phi_e}  \min_{\gamma^e(\cdot)}\mathbb{E}_e[(m-\mathbb{E}_d[m|\gamma^e(m)+w])^2] + \lambda \mathbb{E}_e[x^2]\nn \\
& \geq \liminf_{\phi_d\to\phi_e} \min_{\gamma^e(\cdot)}\mathbb{E}_e[(m-\mathbb{E}_e[m|\gamma^e(m)+w])^2] + \lambda \mathbb{E}_e[x^2] \nn\\
& = \min_{\gamma^e(\cdot)}\mathbb{E}_e[(m-\mathbb{E}_e[m|\gamma^e(m)+w])^2] + \lambda \mathbb{E}_e[x^2],\nn
\end{align}
where the equality follows from the fact that the optimization problem does not depend on $\phi_d(\cdot)$. This proves the lower semi continuity of encoder's cost. 
\end{enumerate}
\end{IEEEproof}

Theorem~\ref{thm:robustness} reveals an interesting duality property of encoder's cost in the sense that if $\phi_e\to\phi_d$, then upper semi continuity holds whereas if $\phi_d\to\phi_e$, then lower semi continuity holds. On the other hand, for mismatched Gaussian priors, both upper semi continuity and lower semi continuity hold, which proves the robustness of the Stackelberg equilibria around the point of identical priors.

\begin{rem}
We note that the property of upper semi continuity also holds when the source is multidimensional. On the other hand, when the source is multidimensional Gaussian from the perspective of both players, then the analysis regarding lower semi continuity in Theorem~\ref{thm:robustness} does not apply since the optimal encoding policy may not be linear for the team theoretic setup with a multidimensional Gaussian source. Nonetheless, if there is an affine policy restriction for the encoder, a similar analysis to that in \eqref{eq:robustnessGaussian} can be carried out to obtain lower semi continuity result for a setup with multidimensional Gaussian priors.  
\end{rem}

We note that in a related work \cite{WuVerdu2012} in this context, continuity property of minimum mean squared error is investigated in a team theoretic setup. In particular, it is shown that minimum mean squared error is continuous for the case with a linear encoder and an additive channel where the noise density is continuous and bounded \cite[Theorem~4]{WuVerdu2012} and the Gaussian density satisfies these properties. In other words, there is essentially no encoding in \cite{WuVerdu2012} other than a scaling factor and the analysis takes only decoding into account from an information transmission perspective. As opposed to \cite{WuVerdu2012}, our work analyzes the scenario with prior mismatch and there is an encoder which may also be nonlinear.

\subsection{Affine Policies may no longer be Optimal for Gaussian Signaling even with Gaussian Subjective Priors}\label{sec:LackOfOptimal}

Here, the subjective probabilistic beliefs of the encoder and the decoder are taken as Gaussian. In particular, the Gaussian source has different mean and variance from the perspectives of the encoder and the decoder; i.e., the source is $M\sim\phi_e(m)=\mathcal{N}(\mu_e,\sigma_e^2)$ and $M\sim\phi_d(m)=\mathcal{N}(\mu_d,\sigma_d^2)$ from encoder's and decoder's perspective, respectively. In addition, the additive noise, which is independent of the source, is modeled by a zero-mean Gaussian random variable; i.e., the noise is $W\sim\mathcal{N}(0,\sigma_W^2)$.

For Gaussian signaling, affine class of policies is an important class of policies due to its desirable optimality property for the classical team theoretic communication setup with identical costs and priors. On the other hand, for the Stackelberg setup, the optimality of affine policies for Gaussian signaling no longer holds due to the presence of subjective probabilistic beliefs of the players, and thus team-theoretic optimality of linear/affine policies are not robust to perturbations. In order to show this result, the following theorem provides examples where nonlinear encoding policies yield better performance for the encoder than the best affine policies.

\begin{thm}\label{thm:StackelbergSoftQuant}
Consider the quadratic signaling games problem with subjective Gaussian priors where there is no affine policy restriction at the encoder. Then, for a soft power constrained or a hard power constrained encoder, it is not necessarily true that an affine policy always gives the Stackelberg equilibrium solution.
\end{thm}
\begin{IEEEproof}
It suffices to provide examples where a nonlinear policy yields a better cost than the best affine policy. First, consider the soft power constrained setup. We provide an example where a quantization policy leads to a cost (for the encoder) which is better then the optimal cost under affine policy restriction. Let $\mu_e=\mu_d=0$, $\sigma_e^2=6.25$, $\sigma_d^2=0.25$, $\sigma_W^2=0.25$ and $\lambda=1.5$. The aim of the encoder is to solve the optimization problem \eqref{eq:JeSoft} after plugging in the best response of the decoder which is given by $u=\mathbb{E}_d[m|\gamma^e(m)+w]$. Under affine policy restriction at the encoder, the best response of decoder also becomes affine. Numerically solving the optimization problem at the encoder under affine policy restriction leads to the optimal encoding policy $\gamma^{*,e}(m)=0.30\, m$ and the corresponding cost of the encoder is given by $J_{\text{affine}}^{*,e}= 6.12$. Now consider an encoding policy in the form of a quantization policy specified by $\gamma^e(m) = \sqrt{P}\,\mathrm{sgn}(m)$ with $P=0.5$. Such a quantization policy is in fact used in the seminal work of Witsenhausen while constructing a nonlinear control policy that outperforms the best linear policy for the considered control system \cite{witsenhausen1968counterexample}. In our setting, the best response of the decoder to such a quantization policy at the encoder can be computed as $\gamma^d(y) = \sqrt{2/\pi}\sigma_d\, \mathrm{tanh}(\sqrt{P}y/\sigma_W^2)$. With this best response of the decoder, we compute the expected cost of the encoder via numerical integration and obtain $J_{\text{quantization}}^e\approx5.90$. This shows that such a quantization policy outperforms the best affine policy. 

Now, consider the hard power constrained case. Let $\mu_e=\mu_d=0$, $\sigma_e^2=1$, $\sigma_d^2=3$, $\sigma_W^2=0.4$ and $\overline{P}=0.1$. In this case, by numerically solving the optimization problem \eqref{eq:JeHard} under affine policy restriction with decoder's best response plugged in yields a non-informative equilibrium. In this case, the cost of the encoder becomes $J_{\text{affine}}^{*,e}=\sigma_e^2=1$. Now consider an encoding policy in the form of a quantization policy specified by $\gamma^e(m) = \sqrt{\overline{P}}\,\mathrm{sgn}(m)$. In a similar manner to the soft power constrained case, by computing the expected cost of the encoder numerically, we obtain $J_{\text{quantization}}^e\approx0.94$. This shows that such a quantization policy outperforms the best affine policy.
\end{IEEEproof}

Theorem~\ref{thm:StackelbergSoftQuant} shows that when the players have different perception on the prior probability of the message, then the optimality of linear policies (which holds under identical priors) may break down. Nonetheless, it is observed through numerical simulations that such a quantization policy outperforms the best affine policy when the players have very different perceptions on the prior probability of the source. 

\subsection{Gaussian Signaling under Affine Policy Restriction} \label{sec:signalingSoft}

In the following theorem, we analyze the Stackelberg equilibrium under the affine policy restriction where there exists a soft power constraint at the encoder. In addition to its simplicity for the Gaussian case, one other motivation for employing affine policies is as follows:

\begin{rem}
Note that in the classical team theoretic setup with identical priors, the optimal solution is attained by affine policies. We also know that the Stackelberg equilibrium concept is used to model scenarios where the decoder trusts the encoder and employs its best response. In such a setting, if the encoder employs a nonlinear policy, then the decoder will be able to learn that its prior is incorrect (as otherwise the encoder must employ an affine policy). As a result, the decoder realizes that the setup is a game setup with mismatched priors (or the decoder may even think that the encoder employs a different cost function). Therefore, the encoder may prefer employing an affine policy rather than a nonlinear policy in order not to damage its credibility. In the case of affine encoding policies, since the decoder may not have the knowledge of power constraint at the encoder, the decoder cannot extract the subjective prior distribution of the encoder using the announced encoding policy. Thus, from an affine policy announcement, the decoder may not realize that its prior is inconsistent in general. 
\end{rem}

\begin{thm}\label{thm:StackelbergSoft}
In the quadratic signaling games with subjective Gaussian priors under soft power constraint, the affine Stackelberg equilibrium is informative if
\begin{align}
\lambda \sigma_e^2 \sigma_W^2<\sigma_d^2(2\sigma_e^2+2(\mu_e-\mu_d)^2-\sigma_d^2) .\label{eq:decreasingFn}
\end{align}
When \eqref{eq:decreasingFn} does not hold, there exists an informative affine Stackelberg equilibrium if the following conditions simultaneously hold:
\begin{align}
3(\sigma_e^2+(\mu_d&-\mu_e)^2)<2\sigma_d^2,\label{eq:concaveFirst} \\
4\lambda \sigma_e^2\sigma_W^2(\sigma_d^2-\sigma_e^2-&(\mu_e-\mu_d)^2) \nn\\ 
&\leq \sigma_d^2(\sigma_e^2+(\mu_e-\mu_d)^2)^2.\label{eq:deltaGreaterThan0}
\end{align}
Otherwise, the affine Stackelberg equilibrium is non-informative.
\end{thm}
\begin{IEEEproof}
See Section~\ref{proof:StackelbergSoft}.
\end{IEEEproof}

\begin{rem}
It is noted that the nature of the affine Stackelberg equilibrium can be non-informative or informative depending on the system parameters. This is because the conditions in \eqref{eq:decreasingFn}-\eqref{eq:deltaGreaterThan0} may or may not hold. For instance, the following are examples of informative and non-informative scenarios:
\begin{enumerate}
\item Let $\sigma_e^2=1$, $\sigma_d^2=4$, $\sigma_W^2=0.25$, $\lambda=2$ and $\mu_e=\mu_d$. In this case, \eqref{eq:concaveFirst} holds whereas \eqref{eq:decreasingFn} and \eqref{eq:deltaGreaterThan0} do not hold, which lead to a non-informative equilibrium. 
\item Let $\sigma_e^2=1$, $\sigma_d^2=4$, $\sigma_W^2=0.25$, $\lambda=1$ and $\mu_e=\mu_d$. As \eqref{eq:concaveFirst} and \eqref{eq:deltaGreaterThan0} are satisfied in this case, the affine Stackelberg equilibrium is informative.
\end{enumerate} 
\end{rem}

\begin{rem}
The conditions in Theorem~\ref{thm:StackelbergSoft} for the informativeness of the affine Stackelberg equilibrium depend on the subjective priors of both players. In particular, Theorem~\ref{thm:StackelbergSoft} shows that the Stackelberg equilibrium solution in general does not coincide with the team theoretic solution when the common prior is the subjective prior of either of the players.
\end{rem}

\begin{rem} \label{rem:inconsistentStackelberg}
When the consistent priors and the zero-mean Gaussian source are assumed; i.e., $\mu_e=\mu_d=0$ and $\sigma_e^2=\sigma_d^2=\sigma_M^2$, then \eqref{eq:decreasingFn} turns into the condition that $\lambda\sigma_W^2<\sigma_M^2$ whereas \eqref{eq:concaveFirst} is always violated. Therefore, we recover the result of \cite[Theorem~4.5]{tacQuadraticSignalingSaritas} (with $b=0$) whose proof indicates that the affine equilibrium is informative if $\lambda< \sigma_M^2/\sigma_W^2$ and non-informative otherwise.
\end{rem}

The analysis in Theorem~\ref{thm:StackelbergSoft} can be carried over to the $N$-stage signaling game: the encoder searches over the affine class to find its optimal policy by anticipating the best response of the decoder, and this would involve an optimization over $N^2+N$ parameters for an $N$-stage problem.

Next, we analyze the Stackelberg equilibrium under the affine policy restriction when there is a hard power constraint at the encoder. Due to prior mismatch, the affine equilibrium solution may be non-informative in contrast with the informative nature of team theoretic solution. The following theorem presents a necessary and sufficient condition for the informativeness of the affine Stackelberg equilibria.

\begin{thm}\label{thm:StackelbergHard}
In the quadratic signaling games with subjective Gaussian priors under hard power constraint, there exists a unique informative affine Stackelberg equilibrium if\footnote{In the case of equality in \eqref{eq:StackelbergHardCond}, it is possible to have a non-informative equilibrium as well and the informative and non-informative equilibria induce the same cost for the encoder in this case.}
\begin{align}
\frac{\sigma_e^2}{\sigma_e^2+(\mu_e-\mu_d)^2}
-\frac{2\sigma_e^2}{\sigma_d^2}
\leq \frac{\overline{P}}{\sigma_W^2} . \label{eq:StackelbergHardCond}
\end{align}
Otherwise, the affine equilibrium is always non-informative. 
\end{thm}

\begin{IEEEproof}
See Section~\ref{proof:StackelbergHard}.
\end{IEEEproof}

When $\mu_e=\mu_d =\mu_M$ and $\sigma_e=\sigma_d =\sigma_M$, i.e., the team theoretic setup, the left hand side of \eqref{eq:StackelbergHardCond} becomes negative leading to a fully informative scenario, as expected. 

\begin{rem}
Unlike the team theoretic solution which is always informative, the affine Stackelberg equilibrium solution is either informative or non-informative depending on the system parameters. 
\end{rem}

Now, we derive the costs at the affine Stackelberg equilibrium in the hard power constrained case to illustrate the effects of subjectivity.  

\begin{thm}\label{thm:StackelbergHardCost}
In the quadratic signaling games with subjective Gaussian priors and hard power constraint, if \eqref{eq:StackelbergHardCond} holds, then the cost of the encoder and decoder at the informative affine Stackelberg equilibrium are
\begin{align}
J_s^{*,e}& = 
\frac{\overline{P}\sigma_d^4\sigma_e^2\sigma_W^2+\sigma_W^4\sigma_e^6+\sigma_W^4\sigma_e^4(\mu_e-\mu_d)^2}
{(\overline{P}\sigma_d^2+\sigma_e^2\sigma_W^2)^2}
, \label{eq:StackelbergHardCostEncoder}\\ 
J_s^{*,d}&= 
\frac{\sigma_e^2\sigma_d^2\sigma_W^2}
{\overline{P}\sigma_d^2+\sigma_e^2\sigma_W^2}
.\label{eq:StackelbergHardCostDecoder}
\end{align}
On the other hand, if \eqref{eq:StackelbergHardCond} does not hold, the affine Stackelberg equilibrium is non-informative with costs $J_s^{*,e}=\sigma_e^2+(\mu_e-\mu_d)^2$ and $J_s^{*,d}= \sigma_d^2$.
\end{thm}

\begin{IEEEproof}
See Section~\ref{proof:StackelbergHardCost}. 
\end{IEEEproof}

It is noted that if both players share a common prior with $\sigma_e=\sigma_d=\sigma_M$ and $\mu_e=\mu_d=\mu_M$ leading to a team theoretic setup, then 
\eqref{eq:StackelbergHardCostEncoder} and \eqref{eq:StackelbergHardCostDecoder} reduces to 
$
J_t^{*,e}= J_t^{*,d}=
\frac{\sigma_M^2\sigma_W^2}
{\overline{P}+\sigma_W^2}.
$

\begin{rem}
Note that one can analyze the Stackelberg equilibrium solutions under affine policy restriction to investigate continuity and robustness properties. On the other hand, we can conclude the continuity result directly from Theorem~\ref{thm:robustness}. In particular, the analysis in the proof of Theorem~\ref{thm:robustness} can be carried out when the optimization problem imposes an additional affine policy restriction. This implies that for the Gaussian prior case, the Stackelberg equilibrium cost of the encoder under affine policy restriction is robust with respect to perturbations around the team setup under either a soft or a hard power constraint. 
\end{rem}

\section{The Non-Cooperative/No-Commitment Setup (Nash Equilibria)} 
In certain communication applications, it may not possible for the encoder and decoder to inform their policies to the other decision maker apriori. In addition, subjective prior distributions of the decision makers may not be known by the other decision maker. In these applications, a Nash theoretic treatment is required to analyze the interaction between the decision makers where there is no hierarchy in the decision making procedure. Furthermore, this type of interaction is used to model scenarios where the decision makers guard themselves against a misleading announcement from the other decision maker, i.e., a decision maker employs a policy other than its announced policy to gain advantage. Motivated by such applications, we investigate affine Nash equilibria when the encoder have subjective probabilistic beliefs on the source distribution where these subjective prior distributions are Gaussian, i.e., the source is $M\sim\phi_e(m)=\mathcal{N}(\mu_e,\sigma_e^2)$ and $M\sim\phi_d(m)=\mathcal{N}(\mu_d,\sigma_d^2)$ from encoder's and decoder's perspective, respectively. 

We first focus on the soft power constrained setup. The following theorem provides a condition under which the affine Nash equilibrium solution is informative or non-informative.

\begin{thm} \label{thm:NashScalarSoft}
In the quadratic signaling games with subjective Gaussian priors and soft power constraint, if $\lambda \geq  \frac{\sigma_d^2}{\sigma_W^2} $, the unique affine equilibrium is non-informative; otherwise, there exists a unique informative affine Nash equilibrium.
\end{thm}
\begin{IEEEproof}
See Section~\ref{proof:NashSoft}.
\end{IEEEproof}

\begin{rem}\label{rem:softPoliciesIndOfEncoder}
From the proof of Theorem~\ref{thm:NashScalarSoft}, it is seen that none of the parameters that specify policies at the equilibrium depend on the subjective priors from the perspective of the encoder since the encoder minimizes its cost for every realization $m$ of source $M$ without considering its distribution.
\end{rem}

\begin{rem}\label{rem:softPoliciesGameTheoretic}
It is seen that the policies for the equilibrium characterized in Theorem~\ref{thm:NashScalarSoft} are the same policies as in the team theoretic setup where both the encoder and decoder take $\phi_d(m)$, i.e., the subjective prior of the decoder in the game theoretic setup, as the source distribution. This is in contrast with the Stackelberg setup where the equilibrium solution does not reduce to a team theoretic solution when either of the player's prior is taken as the common prior in general. 
\end{rem}

Next, we investigate effects of the subjectivity in priors on the equilibrium cost. Due to Remark~\ref{rem:softPoliciesGameTheoretic}, only the objective function of the encoder deviates from its value with the team theoretic solution.

\begin{thm}\label{thm:NashSoftCost}
In the quadratic signaling games with subjective Gaussian priors and soft power constraint, when $\lambda < {\sigma_d^2 \over \sigma_W^2}$, the encoder cost $J_s^{*,e}$ and decoder cost $J_s^{*,d}$ at the Nash equilibrium are
\begin{align}
J_s^{*,e} &= \sqrt{\lambda \sigma_d^2 \sigma_W^2}\left({\sigma_e^2+\sigma_d^2+(\mu_e-\mu_d)^2\over\sigma_d^2}\right)-\lambda \sigma_W^2 \,, \label{eq:NashSoftJeInfo}\\
J_s^{*,d} &= \sqrt{\lambda \sigma_d^2 \sigma_W^2} \,.\label{eq:NashSoftJdInfo}
\end{align}
Otherwise; i.e., if $\lambda \geq   {\sigma_d^2 \over \sigma_W^2}$, the costs are $J_s^{*,e} = \sigma_e^2+(\mu_e-\mu_d)^2$ and $J_s^{*,d} = \sigma_d^2$.
\end{thm}
\begin{IEEEproof}
See Section~\ref{proof:NashSoftCost}.
\end{IEEEproof}

If the priors were equal as $\mu_e=\mu_d=\mu_M$ and $\sigma_e=\sigma_d=\sigma_M$; i.e., the team case with a soft power constraint, then the costs are given by 
$
J_t^{*,e} = 2\sqrt{\lambda \sigma_M^2 \sigma_W^2}-\lambda \sigma_W^2,\text{ } 
J_t^{*,d} = \sqrt{\lambda \sigma_M^2 \sigma_W^2}
$
for $\lambda < {\sigma_M^2 \over \sigma_W^2}$ (i.e., at the informative equilibrium), and 
$
J_t^{*,e} = \sigma_M^2,\text{ } 
J_t^{*,d} = \sigma_M^2 
$
for $\lambda \geq  {\sigma_M^2 \over \sigma_W^2}$ (i.e., at the non-informative equilibrium). 

We now discuss the robustness of the equilibrium with respect to perturbations around the team setup. Since the encoding and decoding policies do not depend on encoder's subjective prior, perturbing encoder's prior with respect to decoder's prior does not lead to a change in the policies at the equilibrium. On the other hand, if we consider perturbation of decoder's prior with respect to encoder's prior, i.e., $\mu_d=\mu_e+\epsilon_\mu$ and $\sigma_d=\sigma_e+\epsilon_\sigma$, the policies at the equilibrium change in a continuous manner with perturbation. These observations imply that the equilibrium is robust to perturbations around the team setup. Notice that even though the nature of the equilibrium may change depending on $\epsilon_\sigma$, the policies are still altered continuously, which ensures robustness of the equilibrium. 

We next generalize our results to the multi-dimensional source setting. Let the source be $\mathbf{M}\sim\mathcal{N}(\boldsymbol{\mu}_e,\Sigma_e)$ and $\mathbf{M}\sim\mathcal{N}(\boldsymbol{\mu}_d,\Sigma_d)$ from encoder's and decoder's perspective, respectively, and let the channel noise be $\mathbf{W} \sim \mathcal{N}(0,\Sigma_{\mathbf{W}})$. Let the cost function of the encoder and the decoder be $c^e(\boldsymbol{m},\boldsymbol{x},\boldsymbol{u}) = \lVert \boldsymbol{m}-\boldsymbol{u}\rVert^2 + \lambda \lVert \boldsymbol{x}\rVert^2$ and $c^d(\boldsymbol{m},\boldsymbol{u}) = \lVert \boldsymbol{m}-\boldsymbol{u}\rVert^2$. Accordingly, the objective function of the encoder and decoder are expressed as $J^e(\gamma^e,\gamma^d) = \mathbb{E}_e[c^e(\boldsymbol{m},\boldsymbol{x},\boldsymbol{u})]$ and $J^d(\gamma^e,\gamma^d) = \mathbb{E}_d[c^d(\boldsymbol{m},\boldsymbol{u})]$, respectively. Before presenting our result for the case when the encoder have subjective probabilistic models, we first restate a related result that appears in \cite[Theorem~5.1]{tacQuadraticSignalingSaritas} for completeness. In this theorem, a multidimensional signaling game setup with identical priors and a biased encoder is considered. The difference with the setup considered in this paper is that we consider a zero biased encoder and the priors are not identical.

\begin{thm}[{{\cite[Theorem~5.1]{tacQuadraticSignalingSaritas}}}]\label{thm:multiConsistentSignal}
Consider the multidimensional signaling setup with identical priors where the cost functions are given by $c^e(\boldsymbol{m},\boldsymbol{x},\boldsymbol{u}) = \lVert \boldsymbol{m}-\boldsymbol{u}-\boldsymbol{b}\rVert^2 + \lambda \lVert \boldsymbol{x}\rVert^2$ and $c^d(\boldsymbol{m},\boldsymbol{u}) = \lVert \boldsymbol{m}-\boldsymbol{u}\rVert^2$. 
\begin{enumerate}[(i)]
\item The encoder (decoder) is affine for an affine decoder (encoder) in a multi-dimensional signaling game when the priors are consistent.
\item For an affine Nash equilibrium, an encoding policy $\gamma^e(\vmm)=A\vmm+\vC$ must satisfy $A= T(A)$ where $T(A)=(FF^T+\lambda I)^{-1}$ and $F=(A\Sigma_dA^T+\Sigma_{\mathbf{W}})^{-1}A\Sigma_d$.  
\item There exists at least one equilibrium. 
\end{enumerate}
\end{thm}

We note that Remark~\ref{rem:softPoliciesGameTheoretic} is valid also for the multidimensional case. In particular, an equilibrium solution is given by the team theoretic solution by taking the subjective prior of the decoder as the common prior. The following result uses this observation together with Theorem~\ref{thm:multiConsistentSignal} by taking $\boldsymbol{b}=\boldsymbol{0}$. Thus, we state the following result without proof.

\begin{cor}\label{cor:multiInconsistentSignal}
Consider the multidimensional signaling setup with inconsistent priors where the cost functions are given by $c^e(\boldsymbol{m},\boldsymbol{x},\boldsymbol{u}) = \lVert \boldsymbol{m}-\boldsymbol{u}\rVert^2 + \lambda \lVert \boldsymbol{x}\rVert^2$ and $c^d(\boldsymbol{m},\boldsymbol{u}) = \lVert \boldsymbol{m}-\boldsymbol{u}\rVert^2$.
\begin{enumerate}[(i)]
\item The encoder (decoder) is affine for an affine decoder (encoder) in a multi-dimensional signaling game when the priors are inconsistent.
\item For an affine Nash equilibrium, an encoding policy $\gamma^e(\vmm)=A\vmm+\vC$ must satisfy $A= T(A)$ where $T(A)=(FF^T+\lambda I)^{-1}$ and $F=(A\Sigma_dA^T+\Sigma_{\mathbf{W}})^{-1}A\Sigma_d$.  
\item There exists at least one equilibrium. 
\end{enumerate}
\end{cor}

As noted in \cite{tacQuadraticSignalingSaritas}, there always exists a non-informative equilibrium. In fact, it is possible to guarantee the existence of an informative equilibrium considering a special case. In particular, \cite[Theorem~5.1]{tacQuadraticSignalingSaritas} focuses on a special case with diagonal covariance matrices to establish the existence of an informative equilibrium under a certain condition, and this result is valid for the case of inconsistent priors analyzed in this manuscript by taking $\boldsymbol{b}=\boldsymbol{0}$ and employing the subjective prior of the decoder as the common prior. 

It is possible to generalize our results to multi-stage setting. In this context, a related result is presented in \cite{SaritasAutomatica2020}. For completeness, we first restate this result in the following.
\begin{thm}[{{\cite[Theorem~11]{SaritasAutomatica2020}}}]\label{thm:multistageConsistentSignal}
Consider the multi-stage signaling game setup where the priors are identical and the source is scalar or multi-dimensional.
\begin{enumerate}[(i)]
\item If the encoder uses affine policies at all stages, then the decoder is affine at all stages.
\item If the decoder uses affine policies at all stages, then the encoder is affine at all stages.
\end{enumerate}
\end{thm}

By using the results of Theorem~\ref{thm:NashScalarSoft}, Corollary~\ref{cor:multiInconsistentSignal} and Theorem~\ref{thm:multistageConsistentSignal}, the following conclusion can be made. Thus, we state the following result without proof.

\begin{cor}
Even if the priors are inconsistent from the perspectives of the encoder and the decoder in the multi-stage signaling game, affine policies constitute an invariant subspace under best response maps for scalar and multi-dimensional sources under Nash equilibria.
\end{cor}

In the remainder of this section, we focus on the case where the encoder has a hard power constraint, i.e., the encoder's objective is given by \eqref{eq:JeHard}. Unlike the soft power constrained case, there always exists an informative Nash equilibrium for the hard power constrained case as stated in the following theorem. Nevertheless, this informative Nash equilibrium is not the same as the team theoretic solution when decoder's subjective prior is the common prior.

\begin{thm}\label{thm:NashHard}
There always exists an informative affine Nash equilibrium in the hard power constrained scalar quadratic signaling game in contrast to the soft power constrained scalar quadratic signaling game.
\end{thm}
\begin{IEEEproof}
See Section~\ref{proof:NashHard}.
\end{IEEEproof}

By examining the proof of Theorem~\ref{thm:NashHard}, it is seen that when $\mu_e=\mu_d=\mu_M$ and $\sigma_e=\sigma_d=\sigma_M$ leading to a team theoretic setup, the encoding and decoding policies become $\gamma^e(m)= \frac{\sqrt{\overline{P}}}{\sigma_M}(m-\mu_M)$ and $\gamma^d(y) = \frac{\sqrt{\overline{P}}\sigma_M}{\overline{P}+\sigma_W^2}$, as expected.  

\begin{rem}
Note that although both the Nash equilibrium solution and the team theoretic solution are always informative, the resulting policies are not the same in general. In other words, the subjectivity in the priors affects the equilibrium solution.
\end{rem}

\begin{rem}
When there is a hard power constraint, the policies at the equilibrium are affected by the priors from the perspective of both players, which is in contrast with the case of soft power constraint. When there is a hard power constraint, it is shown that the encoder equates its average power level (with respect to its own prior) to the maximum possible level at the equilibrium. As a result, the encoding policy at the Nash equilibrium depends also on the priors of the encoder in the hard power constrained case.
\end{rem}

Then, we investigate effects of subjectivity in priors on the equilibrium cost.
\begin{thm}\label{thm:NashHardCost}
In the quadratic signaling games with subjective Gaussian priors and hard power constraint, the encoder cost $J_s^{*,e}$ and decoder cost $J_s^{*,d}$ at the Nash equilibrium are
	\begin{align*}
	J_s^{*,e} &= {  \left(\overline{P}{\sigma_d^4\over(\mu_e-\mu_d)^2+\sigma_e^2}+((\mu_e-\mu_d)^2+\sigma_e^2)\sigma_W^2\right) \sigma_W^2 \over \left(\overline{P} \left({\sigma_d^2\over(\mu_e-\mu_d)^2+\sigma_e^2}\right) + \sigma_W^2\right)^2} \,, \nn\\
	J_s^{*,d} &= {\sigma_d^2 \sigma_W^2 \over \overline{P} \left({\sigma_d^2\over(\mu_e-\mu_d)^2+\sigma_e^2}\right) + \sigma_W^2} \,.
	\end{align*}
\end{thm}
\begin{IEEEproof}
See Section~\ref{proof:NashHardCost}.
\end{IEEEproof}
If the priors were equal as $\mu_e=\mu_d=\mu_M$ and $\sigma_e=\sigma_d=\sigma_M$; i.e., the team case with a hard power constraint, then, $
J_t^{*,e} = J_t^{*,d} = {\sigma_M^2 \sigma_W^2 \over \overline{P} + \sigma_W^2} .
$

When the priors are perturbed around the team setup, unlike the soft power constrained case, the equilibrium is always informative regardless of the perturbation. In addition, when a perturbation is applied, the policies change in a continuous manner, i.e., the affine Nash equilibrium is robust with respect to perturbations around the team setup.

\section{Signaling under Subjective Priors without Additive Noise: Cheap Talk Re-visited} \label{sec:cheap}

In this section, different from the previous communication scenario, there does not exist an additive noise term, i.e., the decoder observes the encoded message directly, and there is no power constraint at the encoder. The setup is depicted in Fig.~\ref{figure:allScheme}-(\subref{figure:noiselessScheme}). We consider encoders and decoders with subjective priors, and to reflect mainly the effects of the subjectivity in the priors we assume that the costs are identical with $c^e(\vmm,\vuu)=c^d(\vmm,\vuu)=\|\vmm-\vuu\|^2$. In contrast with the previously analyzed scenarios, there does not exist a power constraint at the encoder in this case. In fact, the problem can be viewed as a cheap talk game as the cost function of the encoder does not depend on the transmitted message, i.e., transmitting information does not induce a cost for the encoder. For the cheap talk problem, the existence of babbling equilibrium can always be established and in that respect the following is a useful observation, which follows from \cite[Theorem 1]{SignalingGames} and \cite{multiagentSystemsBook}:

\begin{proposition} \label{prop:nonInEq}
A non-informative (babbling) equilibrium always exists for the cheap talk game.
\end{proposition}

Let $\mathbf{M}\sim f_e$ and $\mathbf{M}\sim f_d$ from the perspectives of encoder and decoder, respectively. We assume mutual absolute continuity of $f_e$ and $f_d$; that is, for any Borel set $B$, $f_e(B) = 0 \implies f_d(B)=0$ and $f_d(B) = 0 \implies f_e(B)=0$. This means that even though the encoder and decoder have subjective priors, they believe that the source has the same support, i.e., there is no inconsistency with regard to the support of the source. For instance, this holds when the source has unbounded support with a strictly positive probability density function (e.g., Gaussian and Laplacian distributions) from the perspectives of both players.

\begin{thm}\label{thm:cheap}
\begin{enumerate}[(i)]
\item If the priors are mutually absolutely continuous, there exists a fully informative Nash equilibrium.
\item If the priors are mutually absolutely continuous, there exists a fully informative Stackelberg equilibrium.
\end{enumerate}
\end{thm}
\begin{IEEEproof}
\begin{enumerate}[(i)]
\item Let the encoder and the decoder use fully informative policies; i.e., the encoder transmits every individual message distinctly as $\vxx=\gamma^e(\vmm)=\vmm$, and the decoder takes unique actions for each distinct message it receives as $\vuu=\gamma^d(\vxx)=\vxx$. Then, the cost of the encoder and the decoder is zero almost surely (due to the mutual absolute continuity assumption, the set of $\vmm$ values in the support of both the encoder and the decoder priors has measure $1$ under either the encoder and the decoder prior); and thus $J^e = \mathbb{E}_{f_e}[\|\vmm-\vuu]\|^2]=0$ and $J^d = \mathbb{E}_{f_d}[\|\vmm-\vuu]\|^2]=0$. Since both the encoder and the decoder achieve the minimum possible cost, none of the players deviate from their current choices; i.e., they prefer to stick at the fully informative policies, which implies that there exists a fully informative equilibrium.
\item Under the Stackelberg assumption, the optimal decoder action is $\vuu^*=\gamma^{*,d}(\vxx)=\mathbb{E}_{f_d}[\vmm|\vxx]$. Then, the encoder aims to choose the optimal encoding policy $\gamma^{*,e}(\vmm)=\vxx^*=\underset{\vxx}{\arg\min}\;\mathbb{E}_{f_e}[\|\vmm-\mathbb{E}_{f_d}[\vmm|\vxx]\|^2]$. Thus, for every possible realization of $\vmm$, the encoder can choose $\vxx=\gamma^e(\vmm)$ such that $\vmm=\mathbb{E}_{f_d}[\vmm|\vxx]$, and this is achievable at fully informative equilibria; i.e., $\gamma^{*,e}(\vmm)=\vxx^*=\vmm$. Under this encoding policy and due to the mutual absolute continuity assumption, the optimal encoder cost is zero almost surely, and the optimal decoder policy is $\vuu^*=\gamma^{*,d}(\vxx)=\vxx=\vmm$, which entails a zero decoder cost almost surely. 
\end{enumerate}
\end{IEEEproof}

\begin{rem}\label{rem:cheap}
Under the mutually absolutely continuous priors assumption, the subjectivity in priors does not make a difference; i.e., both the team setup and game setup result in fully informative equilibria.
\end{rem}

\begin{cor}
Theorem~\ref{thm:cheap} and Remark~\ref{rem:cheap} also apply to the multi-stage case; i.e., if the priors are mutually absolutely continuous, there exist fully informative Nash and Stackelberg equilibria in multi-stage and/or multi-dimensional cheap talk as in the team theoretic setup.
\end{cor}

\section{Numerical Examples}

\begin{figure}
\centering
\subfloat[The encoder's prior is the true prior where $\sigma_e^2=1$.]{%
  \includegraphics[width=1.6in]{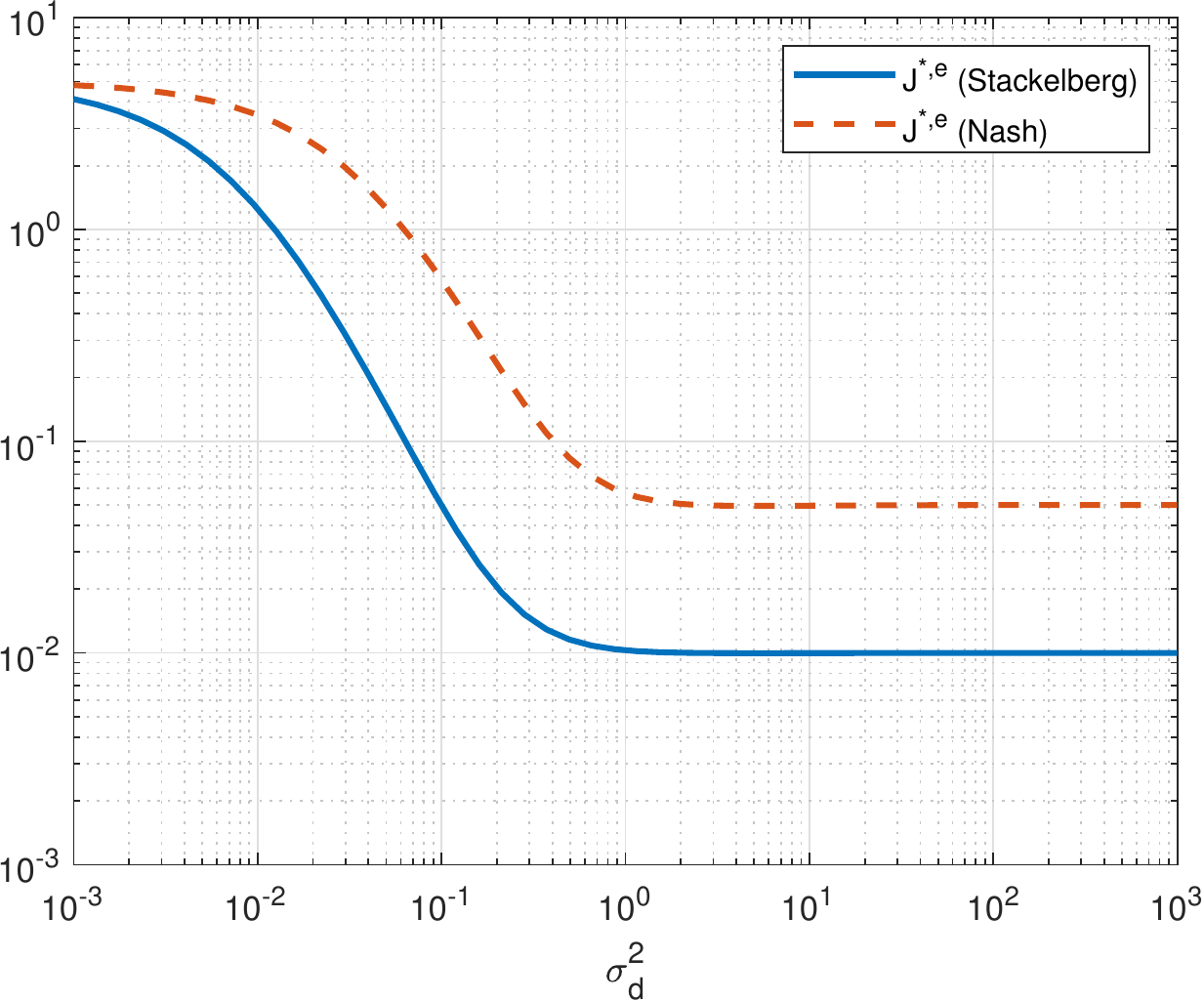}%
  \label{fig:1}%
}\qquad
\subfloat[The decoder's prior is the true prior where $\sigma_d^2=1$.]{%
  \includegraphics[width=1.6in]{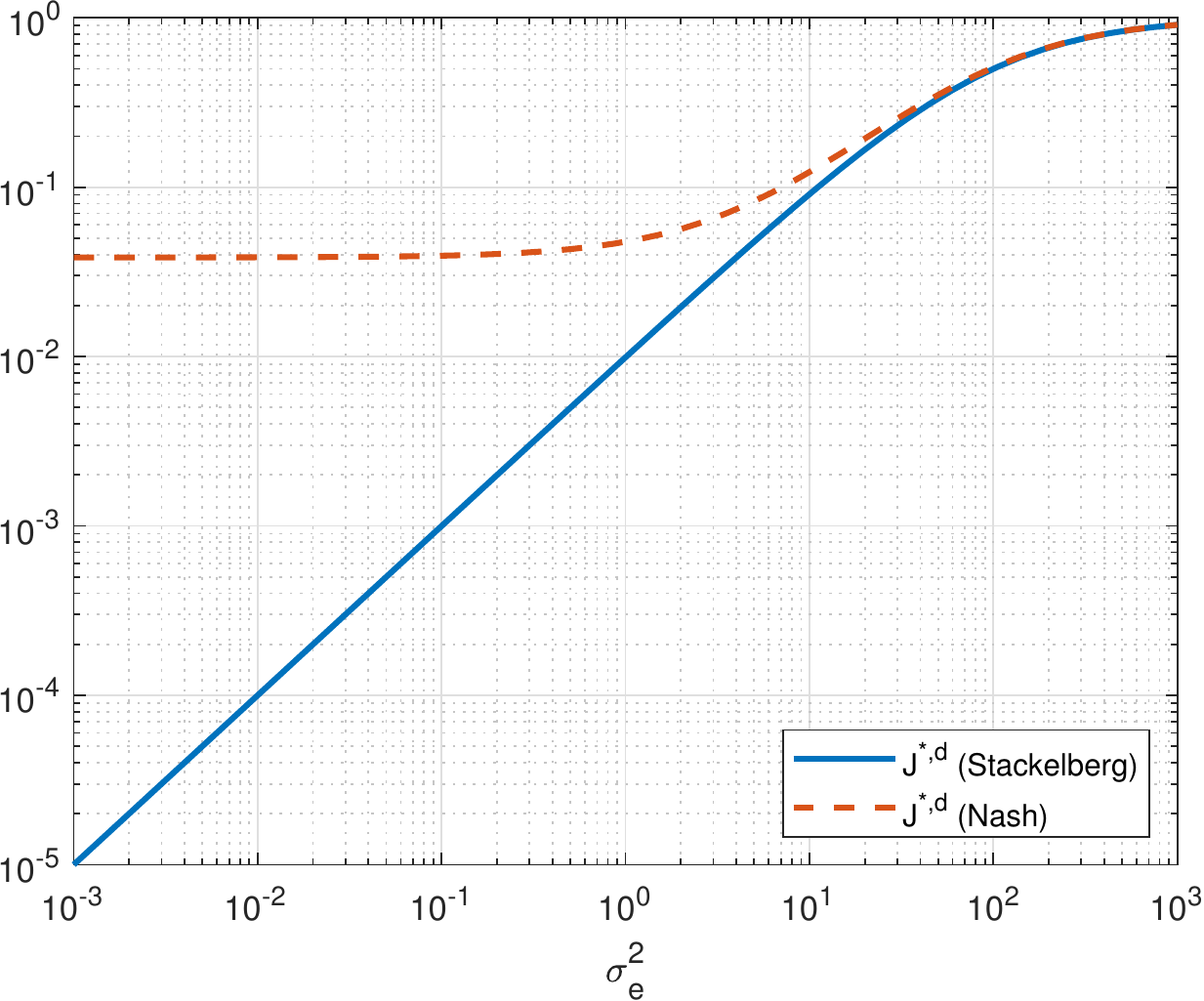}%
  \label{fig:2}%
}

\caption{Attained costs at the unique informative equilibria considering hard power constrained case for the Stackelberg and the Nash setup where $|\mu_e-\mu_d|=2$, $\bar{P}=1$ and $\sigma_W^2=0.01$. The condition \eqref{eq:StackelbergHardCond} for the existence of informative affine Stackelberg equilibria holds for all considered parameter values in the figures.}
\end{figure}

In this section, we provide numerical examples for the Gaussian case. We illustrate the performance values of the unique informative affine Nash and Stackelberg equilibria considering hard power constrained case. In Fig.~\ref{fig:1}, we assume that the prior of the encoder is the true prior, and thus, we plot the induced cost of the encoder at the informative equilibria with respect $\sigma_d^2$. Note that there always exists an informative affine Nash equilibria regardless of the parameter values whereas for the existence of Stackelberg equilibria it is required that \eqref{eq:StackelbergHardCond} holds. For the considered parameter values, this condition always holds. It is observed that for small values of $\sigma_d^2$, the costs are large whereas for large values of $\sigma_d^2$, the costs get smaller for both equilibria. This is intuitive since for small values of $\sigma_d^2$, the decoder has an incorrect belief with high certainty (as $\mu_e\neq \mu_d$) and the encoder cannot persuade the decoder to take an accurate action. On the other hand, for large values of $\sigma_d^2$, the costs get smaller since the conveyed information from the encoder gets more effective. It is also interesting to observe the difference between the costs attained at the informative Stackelberg and the Nash equilibria. This difference arises from the difference in subjective means, and if the subjective means are the same, i.e., $\mu_e=\mu_d$, the costs become the same at the informative Stackelberg and Nash equilibria.

In Fig.~\ref{fig:2}, we now assume that the prior of the decoder is the true prior. Accordingly, we plot the induced cost of the decoder at the informative equilibria with respect $\sigma_e^2$. It is seen that for small values of $\sigma_e^2$, the costs are small whereas for large values of $\sigma_e^2$, the costs get larger for both equilibria. This can be explained by the the following: For large $\sigma_e^2$, the encoder believes that the source has high variance and thus has large average power. Then, in order not to violate the power constraint, the encoder applies a small scaling factor. This effectively reduces the informativeness of the conveyed message since there is an additive noise with a fixed variance. On the other hand, for small $\sigma_e^2$, the encoder applies a large scaling, which in turn reduces the costs. 

\section{Concluding Remarks}

We have investigated Nash and Stackelberg equilibria for quadratic signaling games under subjective/inconsistent priors. We have established qualitative (e.g., full revelation, linearity, informativeness, non-informativeness and robustness around the team setup) and quantitative properties (on linearity or explicit computation) of Nash and Stackelberg equilibria. We have established that the Stackelberg equilibrium cost is upper semi continuous around the point of identical priors and in the particular case of Gaussian priors it is also lower semi continuous, which shows the robustness of the Stackelberg equilibria around the team setup. Moreover, we have shown that for Gaussian signaling, the Stackelberg equilibrium solution is in general nonlinear, which is in contrast with the linearity property of team optimal solution. In addition, we have proven that there exist informative affine Nash and Stackelberg equilibria depending on conditions stated explicitly for the signaling game setup under soft power constraints. Furthermore, we have shown that the Nash equilibrium solution reduces to the team theoretic solution when decoder's prior is the common prior. Moreover, we have established that there always exists a unique affine Nash equilibrium for the signaling game under the hard power constraint as in the team theoretic setup. In addition, we have proven that the affine Stackelberg equilibrium for the signaling game under the hard power constraint is informative or non-informative depending on the system parameters, which is in contrast with the informative nature of the team theoretic solution. Finally, when the source is perceived to admit different probability measures from the perspectives of the encoder and the decoder, under identical cost functions and mutual absolute continuity, we have shown that there exist fully informative Nash and Stackelberg equilibria for the dynamic cheap talk (noiseless case) as in the usual team theoretic setup.

\section{Appendix}

\subsection{Proof of Theorem~\ref{thm:StackelbergSoft}} \label{proof:StackelbergSoft}

Assuming an affine encoder; i.e., $x=\gamma^e(m)=Am+C$, from the previous part, we know that the optimal decoder is $u^*=\gamma^{*,d}(y) = \mathbb{E}_d[m|y] = {A\sigma_d^2\over A^2\sigma_d^2+\sigma_W^2}y+{\sigma_W^2\mu_d-AC\sigma_d^2\over A^2\sigma_d^2+\sigma_W^2}\triangleq Ky+L$, where $y=Am+C+w$. Then, by inserting the best response of the decoder into the objective function of the encoder and after some manipulation, the goal of the encoder becomes 
\begin{align}
&\min_{x = \gamma^e(m)= Am + C}  \mathbb{E}_e \Big[(m-u)^2+\lambda x^2\Big] \nn\\
&\hphantom{phan}=\min_{A,\;C}\; {\sigma_W^4((\mu_e-\mu_d)^2+\sigma_e^2-\sigma_d^2)\over (A^2\sigma_d^2+\sigma_W^2)^2} \nn\\
&\hphantom{phan}\quad+{\sigma_d^2\sigma_W^2\over A^2\sigma_d^2+\sigma_W^2}+\lambda A^2\sigma_e^2+\lambda(A\mu_e+C)^2 \,.
\end{align}
Here, the optimal encoder cost is achieved when $C^*=-A^*\mu_e$, and $A^*$ can be found by solving 
\begin{align}
A^*=\underset{A}{\arg\min} \;& 
\frac{\sigma_W^4(\sigma_e^2+(\mu_e-\mu_d)^2-\sigma_d^2)}{(A^2\sigma_d^2+\sigma_W^2)^2}\nn\\
&\;+{\sigma_d^2\sigma_W^2\over A^2\sigma_d^2+\sigma_W^2}+\lambda A^2\sigma_e^2 \triangleq f(A^2)\,. \label{eq:StackelbergOptimProb}
\end{align}
In order to investigate the optimization problem in \eqref{eq:StackelbergOptimProb}, we need the first and second derivatives of $f(A^2)$ with respect to $A^2$. Taking the first derivative of $f(A^2)$ with respect to $A^2$, we get
\begin{align*}
\frac{df(A^2)}{d(A^2)} 
= -\frac{2\sigma_d^2\sigma_W^4(\sigma_e^2+(\mu_e-\mu_d)^2-\sigma_d^2)}{(A^2\sigma_d^2+\sigma_W^2)^3}\nn\\
- {\sigma_d^4\sigma_W^2\over (A^2\sigma_d^2+\sigma_W^2)^2} + \lambda \sigma_e^2.
\end{align*}
Differentiating further leads to 
\begin{align*}
\frac{d^2f(A^2)}{d(A^2)^2}=
\frac{\sigma_d^4\sigma_W^4\big(
6(\mu_e-\mu_d)^2+6\sigma_e^2-4\sigma_d^2
\big)+2A^2\sigma_d^8\sigma_W^2}{(A^2\sigma_d^2+\sigma_W^2)^4}. 
\end{align*}
It is seen that if \eqref{eq:concaveFirst} holds, the objective function is first concave and then convex when $A^2$ is increased starting from zero. If \eqref{eq:concaveFirst} does not hold, the objective function is always convex. Therefore, a solution to \eqref{eq:StackelbergOptimProb} always exists.

Now, we wish to see if the optimal solution of \eqref{eq:StackelbergOptimProb} leads to $A\neq0$ or not. If the first derivative at $A^2=0$ is negative, it means that there exists $A\neq 0$ which induces a lower cost than the non-informative scenario for the encoder. Therefore, the affine Stackelberg equilibrium is informative under the condition that $\frac{df(A^2)}{d(A^2)}<0 $ when $A=0$, and this condition can be expressed as in \eqref{eq:decreasingFn}. 

If \eqref{eq:decreasingFn} does not hold, then the objective function is non-decreasing at $A^2=0$. Notice that if \eqref{eq:concaveFirst} does not hold, then the objective function becomes strictly convex when $A^2>0$. Since a strictly convex function with a non-negative first derivative at $A^2=0$ is an increasing function for $A^2>0$, the optimal solution becomes non-informative with $A^*=0$ in this case. 

In the remainder of the proof, we consider the remaining scenario when \eqref{eq:concaveFirst} holds and \eqref{eq:decreasingFn} does not hold. In this case, we wish to see if there exists a solution to $f(x)=f(0)$ with a positive $x$. We note that in this remaining case, $f(x)$ is non-decreasing at $x=0$ and strictly concave when $x<0$. Hence, when there exists a solution to $f(x)=f(0)$, this solution must be attained with a positive $x$. Now, we are looking for a solution to the following equation for an informative equilibrium:
\begin{align}
\frac{\sigma_W^4(\sigma_e^2+(\mu_e-\mu_d)^2-\sigma_d^2)}{(A^2\sigma_d^2+\sigma_W^2)^2}
&+\frac{\sigma_d^2\sigma_W^2}{A^2\sigma_d^2+\sigma_W^2}
+\lambda A^2\sigma_e^2  \nn\\
&=\sigma_e^2 + (\mu_e-\mu_d)^2 \label{eq:equalityForInformative}.
\end{align}
Note that \eqref{eq:equalityForInformative} is equivalent to 
\begin{align*}
A^4 \lambda \sigma_d^4\sigma_e^2 
+ A^2 \Big(2\lambda \sigma_e^2\sigma_d^2\sigma_W^2- \sigma_d^4 \big(\sigma_e^2+(\mu_e-\mu_d)^2\big)\Big)\\
+\sigma_d^4\sigma_W^2 +\lambda \sigma_e^2\sigma_W^4
- 2\sigma_d^2\sigma_W^2 \big(\sigma_e^2+(\mu_e-\mu_d)^2\big) =0,
\end{align*}
which is a second order polynomial in $A^2$ whose discriminant is expressed as
$
\Delta \triangleq \sigma_d^8 \big(\sigma_e^2+(\mu_e-\mu_d)^2 \big)^2 
+ 4\lambda \sigma_e^2\sigma_d^2\sigma_W^2 \big(\sigma_e^2+(\mu_e-\mu_d)^2-\sigma_d^2 \big).
$
Hence, when $\Delta\geq 0$ or equivalently \eqref{eq:deltaGreaterThan0} holds, there exists a solution to \eqref{eq:equalityForInformative}. This implies that there exists an informative affine Stackelberg equilibrium in this case.

\subsection{Proof of Theorem~\ref{thm:StackelbergHard}}\label{proof:StackelbergHard}

Assuming an affine encoder; i.e., $x=\gamma^e(m)=Am+C$, from the previous part, we know that the optimal decoder is $u^*=\gamma^{*,d}(y) = \mathbb{E}_d[m|y] = {A\sigma_d^2\over A^2\sigma_d^2+\sigma_W^2}y+{\sigma_W^2\mu_d-AC\sigma_d^2\over A^2\sigma_d^2+\sigma_W^2}\triangleq Ky+L$, where $y=Am+C+w$. Then, after inserting the best response of the decoder, the objective function of the encoder becomes
\begin{align}
&\mathbb{E}_e \Big[(m-u)^2\Big] 
=\mathbb{E}_e \left[(m-AKm-KC-Kw-L)^2\right]\nn\\
&\hphantom{phant} = {A^2\sigma_d^4\sigma_W^2+\sigma_W^4(\sigma_e^2+(\mu_e-\mu_d)^2)\over \left(A^2\sigma_d^2+\sigma_W^2\right)^2}\triangleq f(A^2) \label{eq:fofAsquare}
\end{align}
Thus, the goal of the encoder is to solve the following optimization problem:
\begin{align}
J^{*,e} &= \min_{A,\;C}\; f(A^2) \quad \text{s.t. } (A\mu_e+C)^2+A^2\sigma_e^2\leq \overline{P} \nn.
\end{align}
It is seen that the cost to be minimized is independent of $C$. Therefore, if we set $C=-A\mu_e$, we obtain all feasible values of $A$. Hence, the optimization problem can be solved under the constraint that $A^2\sigma_e^2\leq \overline{P}$. Taking the derivative of the objective function with respect to $A^2$, we get
\begin{align*}
{df(A^2)\over dA^2}
&= {\sigma_d^2\sigma_W^2\left(A^2\sigma_d^2+\sigma_W^2\right)\over\left(A^2\sigma_d^2+\sigma_W^2\right)^4} \nn\\
&\times \Big(\sigma_W^2(\sigma_d^2-2\sigma_e^2-2(\mu_e-\mu_d)^2)-A^2\sigma_d^4\Big).
\end{align*}
Here, if $\sigma_d^2-2\sigma_e^2-2(\mu_e-\mu_d)^2 \leq 0$, then $f(A^2)$ is a decreasing function of $A^2$. Therefore, the minimization can be accomplished by choosing $A^2$ as large as possible satisfying the constraint, which makes the optimal encoding policy $(A^*)^2={\overline{P}\over\sigma_e^2}$ and $C^*=-A^*\mu_e$.

On the other hand, if $\sigma_d^2-2\sigma_e^2-2(\mu_e-\mu_d)^2> 0$, then $f(A^2)$ is first an increasing and then a decreasing function of $A^2$. As the goal is to minimize the cost, the optimal $A^2$ value is obtained by choosing either the largest or the smallest $A^2$ depending on the value of $\overline{P}$. The objective function takes the value of $\sigma_e^2+(\mu_e-\mu_d)^2$ when $A^2=0$. It is necessary to check if the solution of $f(A^2)=\sigma_e^2+(\mu_e-\mu_d)^2$ for nonzero $A^2$ is feasible. In particular, $f(\tilde{A}^2) = (\sigma_e^2+(\mu_e-\mu_d)^2)$ yields $\tilde{A}^2={\sigma_W^2(\sigma_d^2-2\sigma_e^2-2(\mu_e-\mu_d)^2)\over \sigma_d^2(\sigma_e^2+(\mu_e-\mu_d)^2)}$ and such a value is feasible when $\tilde{A}^2\sigma_e^2\leq \overline{P}$. As a result, if \eqref{eq:StackelbergHardCond} holds, then there exists an informative affine equilibrium with $A^*=\pm \sqrt{{\overline{P}\over \sigma_e^2}}$. On the other hand, if \eqref{eq:StackelbergHardCond} does not hold, then the largest possible $A^2$ gives an objective value larger than the objective value at $A=0$, which leads to a non-informative equilibrium. Hence, for a non-informative equilibrium, we need ${\sigma_W^2\sigma_e^2(\sigma_d^2-2\sigma_e^2-2(\mu_e-\mu_d)^2)\over \sigma_d^2(\sigma_e^2+(\mu_e-\mu_d)^2)} >  \overline{P}$ and $\sigma_d^2-2\sigma_e^2-2(\mu_e-\mu_d)^2> 0$ to hold simultaneously and otherwise the affine equilibrium is informative. Since the former condition implies the latter condition, the former condition becomes the only condition which leads to a non-informative equilibrium.

\subsection{Proof of Theorem~\ref{thm:StackelbergHardCost}}\label{proof:StackelbergHardCost}

As shown in Theorem~\ref{thm:StackelbergHard}, an encoding policy $\gamma^{*,e}(m)=Am+C$ with $A= \sqrt{\overline{P}}/\sigma_e$ and $C=-A\mu_e$
leads to an informative affine equilibrium. In response to this encoding policy, the decoder takes its optimal action as $u^*=\gamma^{*,d}(y)=Ky+L=AKm+KC+L+Kw$ with $K=\frac{\sqrt{\overline{P}}\sigma_d^2\sigma_e}{\overline{P}\sigma_d^2+\sigma_e^2\sigma_W^2}$ and $L=\mu_d -K\sqrt{\overline{P}}(\mu_d-\mu_e)/\sigma_e$. By inserting the optimal value of $A$ into \eqref{eq:fofAsquare}, the cost of the encoder becomes as in \eqref{eq:StackelbergHardCostEncoder}. Now, we derive the cost of the decoder at the informative equilibrium. By using best response characterization of the decoder, it can be shown that $KC+L=(1-AK)\mu_d$. Then, the cost of the decoder becomes
\begin{align*}
\mathbb{E}_d[(m-u)^2] 
&= \mathbb{E}_d[(m-K(Am+C+w)-L)^2]\\
&= \mathbb{E}_d[((1-AK)m-(KC+L))^2] + K^2\sigma_W^2\\
&= (1-AK)^2\mathbb{E}_d[(m-\mu_d)^2] + K^2\sigma_W^2\\
&= \frac{\sigma_e^2\sigma_d^2\sigma_W^2}
{\overline{P}\sigma_d^2+\sigma_e^2\sigma_W^2}\,.
\end{align*} 

When \eqref{eq:StackelbergHardCond} does not hold, an encoding policy $\gamma^e(m)=C$ with $C\leq \sqrt{P}$ leads to an affine Stackelberg equilibrium. In this case, the best response of the decoder becomes $\gamma^d(y)=\mathbb{E}_d[m|y]=\mathbb{E}_d[m]=\mu_d$. Then, the costs are given by
\begin{align*}
J_s^{*,e} &= \mathbb{E}_e[(m-u)^2] = \mathbb{E}_e[(m-\mu_d)^2] = \sigma_e^2 + (\mu_e-\mu_d)^2,\\
J_s^{*,d} &= \mathbb{E}_d[(m-u)^2] = \mathbb{E}_d[(m-\mu_d)^2] = \sigma_d^2.
\end{align*}

\subsection{Proof of Theorem~\ref{thm:NashScalarSoft}}\label{proof:NashSoft}

If the encoder is affine; i.e., $x=\gamma^e(m)=Am+C$, then the optimal decoder becomes
\begin{align*}
u^*&=\gamma^{*,d}(y)=\mathbb{E}_d[m|y]=\mathbb{E}_d[m|Am+C+w]\nn\\
&=\mu_d+{A\sigma_d^2\over A^2\sigma_d^2+\sigma_W^2}(y-A\mu_d-C) \nn\\
&= {A\sigma_d^2\over A^2\sigma_d^2+\sigma_W^2}y+{\sigma_W^2\mu_d-AC\sigma_d^2\over A^2\sigma_d^2+\sigma_W^2}.
\end{align*}
Now suppose that the decoder is affine; i.e., $u=\gamma^d(y)=Ky+L$, then the optimal encoder is given by \cite{tacQuadraticSignalingSaritas}
\begin{align*}
x^*=\gamma^{*,e}(m)={K\over K^2+\lambda}m-{KL\over K^2+\lambda}.
\end{align*}

We now wish to see if these optimal sets of policies satisfy a fixed point equation. By combining the optimal policies, we get
\begin{align}
A= \frac{K}{K^2+\lambda}, &\quad K=\frac{A\sigma_d^2}{A^2\sigma_d^2+\sigma_W^2}, \label{eq:softSolutions1}\\
C=\frac{-KL}{K^2+\lambda}, & \quad 
L=\frac{\sigma_W^2\mu_d-AC\sigma_d^2}{A^2\sigma_d^2+\sigma_W^2}. \label{eq:softSolutions2}
\end{align}
Since $1-AK = \frac{\sigma_W^2}{A^2\sigma_d^2+\sigma_W^2}$ and $C=-AL$ from \eqref{eq:softSolutions1} and \eqref{eq:softSolutions2}, it follows that $L=\mu_d$ and $C=\frac{-K\mu_d}{K^2+\lambda}$. Hence, it suffices to find $A$ and $K$ that satisfy \eqref{eq:softSolutions1} for an affine equilibrium. Similar to \cite[Theorem 4.1]{tacQuadraticSignalingSaritas}, we obtain $(K^2+\lambda)^2 \sigma_W^2 = \lambda \sigma_d^2$ from \eqref{eq:softSolutions1} by assuming $A\neq0$. Here, for $\lambda>{\sigma_d^2 \over \sigma_W^2}$, $(K^2+\lambda)^2 \sigma_W^2 = \lambda \sigma_d^2$ cannot be satisfied, thus $A=0$ and the affine equilibrium is non-informative. If $\lambda={\sigma_d^2 \over \sigma_W^2}$ holds, then $(K^2+\lambda)^2 \sigma_W^2 = \lambda \sigma_d^2$ leads to $K=0$ and thus the affine equilibrium is non-informative. Finally, if $\lambda<{\sigma_d^2 \over \sigma_W^2}$ holds, the unique informative affine Nash equilibrium is attained by the encoding and decoding policies $\gamma^e(m)=Am+C$ and $\gamma^d(y) = Ky+L$ with 
\begin{align}
&A = \gamma, \quad C = -\mu_d \gamma, \label{eq:softSolutions3}\\
&K = (\gamma \sigma_d\sqrt{\lambda})/\sigma_W,\quad L = \mu_d \label{eq:softSolutions4}
\end{align}
where $\gamma\triangleq \pm \sqrt{\sqrt{\sigma_W^2\over\lambda\sigma_d^2}-{\sigma_W^2\over\sigma_d^2}} $.

\subsection{Proof of Theorem~\ref{thm:NashSoftCost}}\label{proof:NashSoftCost}

We first consider the informative equilibrium; i.e., the case with $\lambda < {\sigma_d^2 \over \sigma_W^2}$. At the affine Nash equilibrium, the encoder policy is $x^*=\gamma^{*,e}(m)=Am+C$, and the decoder receives $y=x+w=Am+C+w$. Then, the decoder takes its optimal action as $u^*=\gamma^{*,d}(y)=Ky+L=AKm+KC+L+Kw$. From \eqref{eq:softSolutions1}-\eqref{eq:softSolutions4}, it follows that $L+KC=L+K(-AL)=L(1-AK)=\mu_d(1-AK)$ and $1-AK = {\lambda\over K^2+\lambda} =\sqrt{\lambda\sigma_W^2\over\sigma_d^2}$. Now observe the following:
\begin{align*}
\mathbb{E}[(m-u)^2] &= \mathbb{E}[(m-AKm-KC-L-Kw)^2] \nn\\
&= \mathbb{E}[(m(1-AK)-\mu_d(1-AK))^2]+K^2\sigma_W^2 \nn\\
&={\lambda\sigma_W^2\over\sigma_d^2} \mathbb{E}[(m-\mu_d)^2] + \sqrt{\lambda\sigma_d^2\sigma_W^2}-\lambda\sigma_W^2 \,.
\end{align*}
Since $x=Am+C=Am-AL=A(m-\mu_d)$, the cost of the encoder becomes
\begin{align*}
J_s^{*,e} &=  \mathbb{E}_e[(m-u)^2 + \lambda x^2] \nn\\
&= \left({\lambda\sigma_W^2\over\sigma_d^2}+\lambda A^2\right) \mathbb{E}_e[(m-\mu_d)^2] + \sqrt{\lambda\sigma_d^2\sigma_W^2}-\lambda\sigma_W^2 \nn\\
&=  \sqrt{\lambda\sigma_W^2\over\sigma_d^2} \left(\sigma_e^2+(\mu_e-\mu_d)^2\right) + \sqrt{\lambda\sigma_d^2\sigma_W^2}-\lambda\sigma_W^2 \nn\\
&= \sqrt{\lambda\sigma_d^2\sigma_W^2}\left({\sigma_e^2+\sigma_d^2+(\mu_e-\mu_d)^2\over\sigma_d^2}\right) - \lambda\sigma_W^2 \,.
\end{align*}
Similarly, the cost of the decoder is given by
\begin{align*}
&J_s^{*,d} =  \mathbb{E}_d[(m-u)^2] \nn\\
&= {\lambda\sigma_W^2\over\sigma_d^2} \mathbb{E}_d[(m-\mu_d)^2] + \sqrt{\lambda\sigma_d^2\sigma_W^2}-\lambda\sigma_W^2 
= \sqrt{\lambda\sigma_d^2\sigma_W^2} .
\end{align*}

As stated in Theorem~\ref{thm:NashScalarSoft}, the case with $\lambda \geq  {\sigma_d^2 \over \sigma_W^2}$ leads to a non-informative equilibrium with $A=0$, $C=0$, $K=0$ and $L=\mu_d$ from \eqref{eq:softSolutions1} and \eqref{eq:softSolutions2}. Then, $x^*=\gamma^{*,e}(m)=Am+C=0$, $y=x+w=w$, and $u^*=\gamma^{*,d}(y)=Ky+L=\mu_d$ are obtained. Thus, the objectives of the encoder and decoder at the non-informative equilibrium are given by $J_s^{*,e} =  \mathbb{E}_e[(m-u)^2 + \lambda x^2] =\sigma_e^2+(\mu_e-\mu_d)^2 $
and $J_s^{*,d} =  \mathbb{E}_d[(m-u)^2] = \sigma_d^2$.

\subsection{Proof of Theorem~\ref{thm:NashHard}}\label{proof:NashHard}
For an affine encoder; i.e., $x=\gamma^e(m)=Am+C$ which satisfies $\mathbb{E}_e[x^2]=A^2(\mu_e^2+\sigma_e^2)+ 2AC\mu_e+ C^2\leq\overline{P}$, the optimal decoder is affine; namely, 
\begin{align*}
\gamma^{*,d}(y) = {A\sigma_d^2\over A^2\sigma_d^2+\sigma_W^2}y+{\sigma_W^2\mu_d-AC\sigma_d^2\over A^2\sigma_d^2+\sigma_W^2}.
\end{align*}

For an affine decoder; i.e., $u = \gamma^d(y) = Ky + L$, we investigate the optimal encoder as follows: With $y = \gamma^e(m)+w$, it follows that $u=K\gamma^e(m)+Kw+L$. Then, under the hard power constraint $\mathbb{E}_e\left[\left(\gamma^e(m)\right)^2\right]\leq \overline{P}$, the optimal cost of the encoder can be written as
\begin{align}
J^{*,e} &= \min_{x = \gamma^e(m)} \mathbb{E}_e\left[(m-u)^2\right]\nn\\
& = \min_{\gamma^e(m)} \mathbb{E}_e \Big[(m-K\gamma^e(m)-Kw-L)^2\Big]  \nn\\
& = \min_{\gamma^e(m)} \mathbb{E}_e \Big[(m-K\gamma^e(m)-L)^2\Big] +K^2\sigma_W^2 .
\label{eq:encoderOptimization}
\end{align}
For the optimization problem in \eqref{eq:encoderOptimization}, the corresponding Lagrangian function is expressed as
\begin{align}
\mathcal{L}&\left(\gamma^e(m), \nu\right) = \mathbb{E}_e\left[(m-K\gamma^e(m)-L)^2\right] +K^2\sigma_W^2 \nn\\
&\qquad+ \nu\left(\mathbb{E}_e\left[\left(\gamma^e(m)\right)^2\right]-\overline{P}\right) \nn\\
&=\left(K^2+\nu\right)\mathbb{E}_e\left[\left(\gamma^e(m)-{(m-L)K\over K^2+\nu}\right)^2\right] \nn \\
&\qquad+{\nu\over K^2+\nu}\mathbb{E}_e\left[\left(m-L\right)^2\right]+K^2\sigma_W^2-\nu\overline{P}, \label{eq:lagrangianOriginal}
\end{align}
the dual function is given by
\begin{align}
g(\nu) \triangleq \;&\underset{{\gamma^e(m)}}{\inf}\quad \mathcal{L}\left(\gamma^e(m), \nu\right) ,
\label{eq:lagrangianDual}
\end{align}
and the Lagrangian dual problem of \eqref{eq:encoderOptimization} is defined as
\begin{align}
\underset{{\nu}}{\min} \;  g\left(\nu\right) ~
\text{s.t. } \nu \geq 0\,.
\label{eq:lagrangianDualProblem}
\end{align}
Since the optimization problem is convex, the duality gap between the solutions of the primal and the dual problem is zero.

It is observed from \eqref{eq:lagrangianOriginal} that the Lagrangian function $\mathcal{L}\left(\gamma^e(m), \nu\right)$ can be decomposed into
\begin{align}\label{eq:dualDecompLagr}
\mathcal{L}&\left(\gamma^e(m), \nu\right) = \left(K^2+\nu\right)\left(\int_{m\in\mathbb{R}} \mathcal{L}_m\left(\gamma^e(m), \nu\right) p_e(m)\, \mathrm{d}m\right) \nn\\
&\qquad\quad+ {\nu\over K^2+\nu}\mathbb{E}_e\left[\left(m-L\right)^2\right]+K^2\sigma_W^2-\nu\overline{P}\,,
\end{align}
where $\mathcal{L}_m\left(\gamma^e(m), \nu\right) \triangleq\left(\gamma^e(m)-{(m-L)K\over K^2+\nu}\right)^2$. Evidently, the optimal encoder policy that minimizes $\mathcal{L}\left(\gamma^e(m), \nu\right)$ obtained from \eqref{eq:lagrangianDual} should also minimize $\mathcal{L}_m\left(\gamma^e(m), \nu\right)$ for each given value of $m$. This is known as dual decomposition and it facilitates the decomposition of the dual problem into sub-optimization problems which are coupled only through $m$. More explicitly, we need the compute
\begin{gather}\label{eq:dualDecompProb}
\underset{{\gamma^e(m)}}{\min} \mathcal{L}_m\left(\gamma^e(m), \nu\right) = \underset{{\gamma^e(m)}}{\min} \left(\gamma^e(m)-{(m-L)K\over K^2+\nu}\right)^2
\end{gather}
for each value of $m\in\mathbb{R}$. 

The Karush-Kuhn-Tucker (KKT) conditions can be obtained for the optimization problem in \eqref{eq:encoderOptimization} as follows:
\begin{align}
{\partial \mathcal{L}_m\left(\gamma^e(m), \nu\right)\over\partial \left(\gamma^e(m)\right)} &= 0 \,, \label{eq:kkt1}\\
\nu \left(\mathbb{E}_e\left[\left(\gamma^e(m)\right)^2\right] - \overline{P}\right) &= 0 \,,
\label{eq:kkt2} \\
\nu &\geq 0 \,,
\label{eq:kkt3} \\
\mathbb{E}_e\left[\left(\gamma^e(m)\right)^2\right] - \overline{P} &\leq 0 \,.
\label{eq:kkt4}
\end{align}

From \eqref{eq:kkt1}, the optimal encoder policy is $\gamma^e(m)={K\over K^2+\nu}m-{KL\over K^2+\nu}$. By \eqref{eq:kkt2}, we must have either $\nu=0$ or $\mathbb{E}_e\left[\left(\gamma^e(m)\right)^2\right]=\overline{P}$. If $\nu=0$, for an informative affine equilibrium, $K= {A \sigma_d^2 \over A^2 \sigma_d^2 + \sigma_W^2}$ and $A={K\over K^2+\nu}={1\over K}$ must be satisfied simultaneously, which is not possible. Thus, we must investigate the case of $\mathbb{E}_e\left[\left(\gamma^e(m)\right)^2\right]=\overline{P}$ with $\nu>0$ to obtain the conditions for an informative affine equilibrium. More specifically, we need
\begin{align}
\overline{P}&=\mathbb{E}_e\left[\left(\gamma^e(m)\right)^2\right] = \mathbb{E}_e\left[\left({K\over K^2+\nu}m-{KL\over K^2+\nu}\right)^2\right]\nn\\
&={K^2\over (K^2+\nu)^2}\left(\mu_e^2+\sigma_e^2-2L\mu_e+L^2\right) \label{eq:NashHardPowerEqual}
\end{align}
with $\nu>0$. At the equilibrium, we have
\begin{align}
A &= {K\over K^2+\nu}\,,\quad K = {A \sigma_d^2 \over A^2 \sigma_d^2 + \sigma_W^2} \,, \label{eq:NashHardFixedPointAK}\\
C &= -{KL\over K^2+\nu}\,,\quad L = {\sigma_W^2\mu_d-AC\sigma_d^2\over A^2\sigma_d^2+\sigma_W^2}\,.\label{eq:NashHardFixedPointCL}
\end{align}
It is seen that \eqref{eq:NashHardFixedPointAK} and \eqref{eq:NashHardFixedPointCL} lead to $L=\mu_d$. From \eqref{eq:NashHardPowerEqual}-\eqref{eq:NashHardFixedPointCL} and $L=\mu_d$, it follows that $A =  \gamma $, $C=-\mu_d \gamma $ and $K = \frac{\gamma\sigma_d^2}{\gamma^2\sigma_d^2+\sigma_W^2} $ where $\gamma \triangleq \pm \sqrt{\overline{P}\over(\mu_e-\mu_d)^2+\sigma_e^2}$. Finally, we need to ensure $\nu>0$. Observe that \eqref{eq:NashHardPowerEqual} and $L=\mu_d$ lead to 
\begin{align*}
\nu 
= \sqrt{K^2\left(\sigma_e^2+(\mu_e^2-\mu_d)^2\right)\over\overline{P}}-K^2 
= \sqrt{\frac{K^2}{A^2}} - K^2 
>0,
\end{align*}
where the second equality follows from $A=\gamma$ and the inequality is due to $AK<1$. This shows that there always exists an informative affine Nash equilibrium.

\subsection{Proof of Theorem~\ref{thm:NashHardCost}}\label{proof:NashHardCost}
At the affine Nash equilibrium, the encoder policy is $x^*=\gamma^{*,e}(m)=Am+C$, and the decoder receives $y=x+w=Am+C+w$. Then, the decoder takes its optimal action as $u^*=\gamma^{*,d}(y)=Ky+L=AKm+KC+L+Kw$. Since $L+KC=L+K(-AL)=L(1-AK)=\mu_d(1-AK)$ and $1-AK = 1-A{A \sigma_d^2 \over A^2 \sigma_d^2 + \sigma_W^2}={\sigma_W^2 \over A^2 \sigma_d^2 + \sigma_W^2}$, the following holds:
	\begin{align*}
	\mathbb{E}&[(m-u)^2] = \mathbb{E}[(m-AKm-KC-L-Kw)^2] \nn\\
	&= \mathbb{E}[(m(1-AK)-\mu_d(1-AK))^2]+K^2\sigma_W^2 \nn\\
	&={\sigma_W^4 \over \left(A^2 \sigma_d^2 + \sigma_W^2\right)^2} \mathbb{E}[(m-\mu_d)^2] + {A^2 \sigma_d^4 \over \left(A^2 \sigma_d^2 + \sigma_W^2\right)^2}\sigma_W^2 \,.
	\end{align*}
	Then, the encoder cost is
	\begin{align*}
	J_s^{*,e} &=  \mathbb{E}_e[(m-u)^2] \nn\\
	&= {\sigma_W^4 \over \left(A^2 \sigma_d^2 + \sigma_W^2\right)^2} \mathbb{E}_e[(m-\mu_d)^2] + {A^2 \sigma_d^4 \over \left(A^2 \sigma_d^2 + \sigma_W^2\right)^2}\sigma_W^2 \nn\\
	&= {\sigma_W^4\left(\sigma_e^2+(\mu_e-\mu_d)^2\right) \over \left(A^2 \sigma_d^2 + \sigma_W^2\right)^2}  + {A^2 \sigma_d^4 \over \left(A^2 \sigma_d^2 + \sigma_W^2\right)^2}\sigma_W^2 \nn\\
	&=  {\sigma_W^2 \left((\sigma_e^2+(\mu_e-\mu_d)^2)\sigma_W^2+{\overline{P}\over(\mu_e-\mu_d)^2+\sigma_e^2}\sigma_d^4\right) \over \left({\overline{P}\over(\mu_e-\mu_d)^2+\sigma_e^2} \sigma_d^2 + \sigma_W^2\right)^2} \,,
	\end{align*}
	whereas the decoder cost is
	\begin{align*}
	J_s^{*,d} &=  \mathbb{E}_d[(m-u)^2] \nn\\
	&= {\sigma_W^4 \over \left(A^2 \sigma_d^2 + \sigma_W^2\right)^2} \mathbb{E}_d[(m-\mu_d)^2] + {A^2 \sigma_d^4 \over \left(A^2 \sigma_d^2 + \sigma_W^2\right)^2}\sigma_W^2 \nn\\
	&= {\sigma_d^2 \sigma_W^2 \over A^2 \sigma_d^2 + \sigma_W^2} 
	={\sigma_d^2 \sigma_W^2 \over \overline{P} \left({\sigma_d^2\over(\mu_e-\mu_d)^2+\sigma_e^2}\right) + \sigma_W^2}\,.
	\end{align*}

\bibliographystyle{IEEEtran}
\bibliography{subjectiveBibliography}

\begin{thebibliography}{10}
\providecommand{\url}[1]{#1}
\csname url@samestyle\endcsname
\providecommand{\newblock}{\relax}
\providecommand{\bibinfo}[2]{#2}
\providecommand{\BIBentrySTDinterwordspacing}{\spaceskip=0pt\relax}
\providecommand{\BIBentryALTinterwordstretchfactor}{4}
\providecommand{\BIBentryALTinterwordspacing}{\spaceskip=\fontdimen2\font plus
\BIBentryALTinterwordstretchfactor\fontdimen3\font minus
  \fontdimen4\font\relax}
\providecommand{\BIBforeignlanguage}[2]{{%
\expandafter\ifx\csname l@#1\endcsname\relax
\typeout{** WARNING: IEEEtran.bst: No hyphenation pattern has been}%
\typeout{** loaded for the language `#1'. Using the pattern for}%
\typeout{** the default language instead.}%
\else
\language=\csname l@#1\endcsname
\fi
#2}}
\providecommand{\BIBdecl}{\relax}
\BIBdecl

\bibitem{WitsenhausenIntrinsic}
H.~S. Witsenhausen, ``The intrinsic model for discrete stochastic control: Some
  open problems,'' in \emph{Control Theory, Numerical Methods and Computer
  Systems Modelling}, A.~Bensoussan and J.~L. Lions, Eds.\hskip 1em plus 0.5em
  minus 0.4em\relax Berlin, Heidelberg: Springer Berlin Heidelberg, 1975, pp.
  322--335.

\bibitem{basols99}
T.~Ba\c{s}ar and G.~Olsder, \emph{Dynamic Noncooperative Game Theory}.\hskip
  1em plus 0.5em minus 0.4em\relax Philadelphia, PA: SIAM Classics in Applied
  Mathematics, 1999.

\bibitem{witsenhausen1968counterexample}
H.~S. Witsenhausen, ``A counterexample in stochastic optimum control,''
  \emph{SIAM J. Control}, vol.~6, no.~1, pp. 131--147, 1968.

\bibitem{BasarCDC2008}
T.~Ba\c{s}ar, ``Variations on the theme of the {W}itsenhausen counterexample,''
  in \emph{Proc. IEEE Conf. Decis. Control (CDC)}, 2008, pp. 1614--1619.

\bibitem{CoverThomasBook}
T.~M. Cover and J.~A. Thomas, \emph{Elements of Information Theory}.\hskip 1em
  plus 0.5em minus 0.4em\relax New Jersey: John Wiley \& Sons, 2006.

\bibitem{YukselBasarBook}
S.~Y\"uksel and T.~Ba\c{s}ar, \emph{Stochastic Networked Control Systems:
  Stabilization and Optimization under Information Constraints}.\hskip 1em plus
  0.5em minus 0.4em\relax Boston, MA: Birkh\"auser, 2013.

\bibitem{SignalingGames}
V.~P. Crawford and J.~Sobel, ``Strategic information transmission,''
  \emph{Econometrica}, vol.~50, pp. 1431--1451, 1982.

\bibitem{bayesianPersuasion}
E.~Kamenica and M.~Gentzkow, ``Bayesian persuasion,'' \emph{American Economic
  Review}, vol. 101, no.~6, pp. 2590--2615, Oct. 2011.

\bibitem{EAkyolProcIEEE2017}
E.~{Akyol}, C.~{Langbort}, and T.~Ba\c{s}ar, ``Information-theoretic approach
  to strategic communication as a hierarchical game,'' \emph{Proc. IEEE}, vol.
  105, no.~2, pp. 205--218, Feb. 2017.

\bibitem{tacQuadraticSignalingSaritas}
S.~Sar{\i}ta\c{s}, S.~Y{\"{u}}ksel, and S.~Gezici, ``Quadratic
  multi-dimensional signaling games and affine equilibria,'' \emph{IEEE Trans.
  Autom. Control}, vol.~62, no.~2, pp. 605--619, Feb. 2017.

\bibitem{EstStrategicSensorsFarokhi2017}
F.~Farokhi, A.~M.~H. Teixeira, and C.~Langbort, ``Estimation with strategic
  sensors,'' \emph{IEEE Trans. Autom. Control}, vol.~62, no.~2, pp. 724--739,
  Feb. 2017.

\bibitem{Tamura2018}
\BIBentryALTinterwordspacing
W.~Tamura, ``Bayesian persuasion with quadratic preferences.'' [Online].
  Available: \url{https://ssrn.com/abstract=1987877}
\BIBentrySTDinterwordspacing

\bibitem{SaritasAutomatica2020}
S.~Sar{\i}ta{\c{s}}, S.~Y{\"u}ksel, and S.~Gezici, ``Dynamic signaling games
  with quadratic criteria under {N}ash and {S}tackelberg equilibria,''
  \emph{Automatica}, vol. 115, p. 108883, 2020.

\bibitem{MOSayinAutomatica2019}
M.~O. Sayin, E.~Akyol, and T.~Ba\c{s}ar, ``Hierarchical multistage {G}aussian
  signaling games in noncooperative communication and control systems,''
  \emph{Automatica}, vol. 107, pp. 9--20, 2019.

\bibitem{TreustPersuasion2019}
M.~L. Treust and T.~Tomala, ``Persuasion with limited communication capacity,''
  \emph{Journal of Economic Theory}, vol. 184, p. 104940, 2019.

\bibitem{treust2018persuasion}
\BIBentryALTinterwordspacing
------, ``Information-theoretic limits of strategic communication,'' 2018.
  [Online]. Available: \url{http://arxiv.org/abs/1807.05147}
\BIBentrySTDinterwordspacing

\bibitem{ProspectBasar2018}
V.~S.~S. {Nadendla}, C.~{Langbort}, and T.~{Başar}, ``Effects of subjective
  biases on strategic information transmission,'' \emph{IEEE Trans. Commun.},
  vol.~66, no.~12, pp. 6040--6049, Dec. 2018.

\bibitem{SaritasISIT2019}
S.~Sar{\i}ta{\c{s}}, P.~Furrer, S.~Gezici, T.~Linder, and S.~Y{\"u}ksel, ``On
  the number of bins in equilibria for signaling games,'' in \emph{Proc. IEEE
  Int. Symp. Inf. Theory (ISIT)}, 2019, pp. 972--976.

\bibitem{VoraCDC2020}
A.~S. {Vora} and A.~A. {Kulkarni}, ``Information extraction from a strategic
  sender over a noisy channel,'' in \emph{Proc. IEEE Conf. Decis. Control
  (CDC)}, 2020, pp. 354--359.

\bibitem{VoraISIT2020}
------, ``Achievable rates for strategic communication,'' in \emph{Proc. IEEE
  Int. Symp. Inf. Theory (ISIT)}, 2020, pp. 1379--1384.

\bibitem{bayesianPersuasionHetPriors}
R.~Alonso and O.~Câmara, ``Bayesian persuasion with heterogeneous priors,''
  \emph{Journal of Economic Theory}, vol. 165, pp. 672 -- 706, 2016.

\bibitem{CGokenTWC2010}
\c{C}. {G\"{o}ken}, S.~{Gezici}, and O.~{Ar{\i}kan}, ``Optimal stochastic
  signaling for power-constrained binary communications systems,'' \emph{IEEE
  Trans. Wireless Commun.}, vol.~9, no.~12, pp. 3650--3661, 2010.

\bibitem{BasTAC85}
T.~Ba\c{s}ar, ``An equilibrium theory for multiperson decision making with
  multiple probabilistic models,'' \emph{IEEE Trans. Autom. Control}, vol.~30,
  no.~2, pp. 118--132, Feb. 1985.

\bibitem{TeneketzisVaraiya88}
D.~Teneketzis and P.~Varaiya, ``Consensus in distributed estimation,'' in
  \emph{Advances in Statistical Signal Processing}, H.~V. Poor, Ed.\hskip 1em
  plus 0.5em minus 0.4em\relax Greenwich: JAI Press, 1988, ch.~10, pp.
  361--386.

\bibitem{CastanonTeneketzis88}
D.~A. Castanon and D.~Teneketzis, ``Further results on the asymptotic agreement
  problem,'' \emph{IEEE Trans. Autom. Control}, vol.~33, no.~6, pp. 515--523,
  June 1988.

\bibitem{SaritasTSP2019}
S.~{Sarıtaş}, S.~{Gezici}, and S.~{Yüksel}, ``Hypothesis testing under
  subjective priors and costs as a signaling game,'' \emph{IEEE Trans. Signal
  Process.}, vol.~67, no.~19, pp. 5169--5183, Oct. 2019.

\bibitem{mismatchedEstimation}
C.~D. Richmond and L.~L. Horowitz, ``Parameter bounds on estimation accuracy
  under model misspecification,'' \emph{IEEE Trans. Signal Process.}, vol.~63,
  no.~9, pp. 2263--2278, May 2015.

\bibitem{estimationRobustnessMismatch}
R.~M. Dufour and E.~L. Miller, ``Statistical estimation with 1/f-type prior
  models: robustness to mismatch and efficient model determination,'' in
  \emph{Proc. IEEE Int. Conf. Acoust. Speech Signal Process. (ICASSP)}, vol.~5,
  May 1996, pp. 2491--2494 vol. 5.

\bibitem{mismatchSurvey}
S.~Fortunati, F.~Gini, M.~S. Greco, and C.~D. Richmond, ``Performance bounds
  for parameter estimation under misspecified models: Fundamental findings and
  applications,'' \emph{IEEE Signal Process. Mag.}, vol.~34, no.~6, pp.
  142--157, Nov. 2017.

\bibitem{Gray1975}
R.~{Gray} and L.~{Davisson}, ``Quantizer mismatch,'' \emph{IEEE Trans.
  Commun.}, vol.~23, no.~4, pp. 439--443, 1975.

\bibitem{MismatchedMMSEISIT2012}
I.~{Esnaola}, A.~M. {Tulino}, and H.~V. {Poor}, ``Mismatched {MMSE} estimation
  of multivariate {G}aussian sources,'' in \emph{Proc. IEEE Int. Symp. Inf.
  Theory (ISIT)}, 2012, pp. 716--720.

\bibitem{MismatchGuessworkITW2019}
S.~{Salamatian}, L.~{Liu}, A.~{Beirami}, and M.~{Médard}, ``Mismatched
  guesswork and one-to-one codes,'' in \emph{Proc. IEEE Inf. Theory Workshop
  (ITW)}, 2019, pp. 1--5.

\bibitem{JubaCompression2011}
B.~Juba, A.~T. Kalai, S.~Khanna, and M.~Sudan, ``Compression without a common
  prior: An information-theoretic justification for ambiguity in language,'' in
  \emph{Innovations in Computer Science}, 2011, p. 79–86.

\bibitem{BravermanIT2019}
M.~{Braverman} and B.~{Juba}, ``The price of uncertain priors in source
  coding,'' \emph{IEEE Trans. Inf. Theory}, vol.~65, no.~2, pp. 1165--1171,
  2019.

\bibitem{GeziciProspect2018}
S.~{Gezici} and P.~K. {Varshney}, ``On the optimality of likelihood ratio test
  for prospect theory-based binary hypothesis testing,'' \emph{IEEE Signal
  Process. Lett.}, vol.~25, no.~12, pp. 1845--1849, 2018.

\bibitem{ProspectVarshney2020}
B.~{Geng}, S.~{Brahma}, T.~{Wimalajeewa}, P.~K. {Varshney}, and
  M.~{Rangaswamy}, ``Prospect theoretic utility based human decision making in
  multi-agent systems,'' \emph{IEEE Trans. Signal Process.}, vol.~68, pp.
  1091--1104, 2020.

\bibitem{WuVerdu2012}
Y.~{Wu} and S.~{Verdu}, ``Functional properties of minimum mean-square error
  and mutual information,'' \emph{IEEE Trans. Inf. Theory}, vol.~58, no.~3, pp.
  1289--1301, 2012.

\bibitem{ADKara2019SIAM}
A.~D. Kara and S.~Y{\"u}ksel, ``Robustness to incorrect priors in partially
  observed stochastic control,'' \emph{SIAM J. Control Optim.}, vol.~57, no.~3,
  pp. 1929--1964, 2019.

\bibitem{villani2009optimal}
C.~Villani, \emph{Optimal transport, old and new}.\hskip 1em plus 0.5em minus
  0.4em\relax Berlin Heidelberg: Springer-Verlag, 2009.

\bibitem{multiagentSystemsBook}
Y.~Shoham and K.~Leyton-Brown, \emph{Multiagent Systems: Algorithmic,
  Game-Theoretic, and Logical Foundations}.\hskip 1em plus 0.5em minus
  0.4em\relax New York, NY, USA: Cambridge University Press, 2009.

\end{thebibliography}

\end{document}